\documentclass[natbib]{svjour3}                     
\smartqed  
\usepackage{graphicx}
\usepackage{mathptmx}      

 \journalname{Space Science Reviews}
\newcommand{\aap}{{Astron. Astrophys.}}
\newcommand{\apj}{{Astrophys. J.}}
\newcommand{\apjl}{{Astrophys. J.}}
\newcommand{\jgr}{{J. Geophys. Res.}}
\newcommand{\mnras}{{Mon. Not. R. Astron. Soc.}}

\newcommand{\solphys}{{Solar Phys.}}
\newcommand{\pasj}{{Pub. Astron. Soc. Japan}}
\newcommand{\nat}{{Nature}}

\newcommand{\mut}{\tilde{\mu}}
\newcommand{\va}{v_{\mathrm{A}}}

\newcommand{\pd}{\partial}
\newcommand{\td}{\tau_{\mathrm{d}}}
\newcommand{\tdp}{\tau_{\mathrm{d}} / P}

\newcommand{\ez}{\mbox{$\hat{{\bf e}}_{\rm z}$}}
\newcommand{\mutilde}{\tilde{\mu}}

\newcommand{\etac}{\eta_{\rm C}}
\newcommand{\etah}{\eta_{\rm H}}

\newcommand{\etactf}{\tilde{\eta}_{{\rm C}f}}
\newcommand{\etactc}{\tilde{\eta}_{{\rm C}c}}
\newcommand{\etactcf}{\tilde{\eta}_{{\rm C}c,f}}

\begin{document}

\title{Damping mechanisms for oscillations in solar prominences}
\titlerunning{Damping of prominence oscillations}        

\author{I\~nigo Arregui         \and
        Jos\'e Luis Ballester 
}

\authorrunning{I. Arregui \& J.L. Ballester}  

\institute{I. Arregui \& J.L. Ballester \at
              Departament de F\'{\i}sica, Universitat de les Illes Balears\\ E-07122 Palma de Mallorca, Spain \\
              Tel.: +34-971-172391\\
              Fax: +34-971-173426\\
              \email{inigo.arregui@uib.es, joseluis.ballester@uib.es}  
}

\date{Received: date / Accepted: date}

\maketitle

\begin{abstract}
Small amplitude oscillations are a commonly observed feature in prominences/fi\-la\-ments. These oscillations appear to be of local nature, are associated to the fine structure of prominence plasmas,  and simultaneous flows and counterflows are also present. The existing observational evidence reveals that small amplitude oscillations, after excited, are damped in short spatial and temporal scales by some as yet not well determined physical mechanism(s). Commonly, these oscillations have been interpreted in terms of linear magnetohydrodynamic (MHD) waves, and this paper reviews the theoretical damping mechanisms that have been recently put forward in order to explain the observed attenuation scales. These mechanisms include thermal effects, through non-adiabatic processes, mass flows, resonant damping in non-uniform media, and partial ionization effects.  The relevance of each mechanism  is assessed by comparing the spatial and time scales produced by each of them with those obtained from observations. Also, the application of the latest theoretical results to perform prominence seismology is discussed, aiming to determine physical parameters in prominence plasmas that are difficult to measure by direct means.

\end{abstract}

\section{Introduction}
\label{intro}

Quiescent solar filaments are clouds of cool and dense plasma suspended against gravity by forces though to be of magnetic origin. They form along the inversion polarity line in or between the weak remnants of active regions. Early filament observations already revealed that their fine structure is apparently composed by many horizontal and thin dark threads \citep{Engvold98}. More recent high-resolution H$_\alpha$ observations obtained with the Swedish Solar Telescope (SST) in La Palma \citep{Lin05} and the Dutch Open Telescope (DOT) in Tenerife \citep{HA06}  have allowed to observe this fine structure with much greater detail (see \citealt{Lin10}, for a review).  The measured average width of resolved thin threads is about $0.3$ arc.sec ($\sim$ $210$  km) while their length is between $5$ and $40$ arc.sec ($\sim$ 3500 - 28000  km). The fine threads of solar filaments seem to be partially filled with cold plasma \citep{Lin05}, typically two orders of magnitude denser and cooler than the surrounding corona, and it is generally assumed that they outline their magnetic flux tubes \citep{Engvold98,Linthesis,Lin05,Engvold08,Martin08,Lin08}. This idea is strongly supported by observations which suggest that they are inclined with respect to the filament long axis in a similar way to what has been found for the magnetic field (\citealt{Leroy80,Bommier94,BL98}).  On the opposite, \cite{HA06} suggest that these dark horizontal filament fibrils are a projection effect.  According to this view, many magnetic field dips of rather small vertical extension, but filled with cool plasma, would be aligned in the vertical direction and the projection against the disk would produce the impression of a horizontal thread.

Oscillations in prominences and filaments are a commonly observed phenomenon. They are usually classified, in terms of their amplitude, as small and large amplitude oscillations \citep{OB02}. This paper is concerned with small amplitude oscillations. Large amplitude oscillations have been recently reviewed by \cite{Tripathi09}.  It is well established that these small amplitude periodic changes are of local nature.  The detected peak velocity ranges from the noise level (down to 0.1 km~s$^{-1}$ in some cases) to 2--3~km~s$^{-1}$, although larger values have also been reported \citep{BashkirtsevMashnich84,Molowny-Horas99}.  Two-dimensional observations of filaments \citep{YiEngvold91,Yi91} revealed that individual fibrils or groups of fibrils may oscillate independently with their own periods, which range between 3 and 20 minutes. More recently, \cite{Linthesis} reports spatially coherent oscillations found over slices of a polar crown filament covering an area of $1.4 \times 54$ arc.sec$^2$  with,  among other, a significant periodicity at 26 minutes, strongly damped after 4 periods.  Furthermore, \cite{Lin07} have shown evidence about traveling waves along a number of filament threads with an average phase velocity of $12$  km s$^{-1}$, a wavelength of $4''$ ($\sim 2800$ \ km), and
oscillatory periods of the individual threads that vary from $3$ to $9$ minutes. Oscillatory events
have been reported from both ground-based observations \citep{Terradas02,Linthesis}, as well as observations from instruments onboard space-crafts, such as SoHO \citep{Blanco99,Regnier01,Pouget06} and Hinode \citep{Okamoto07,Ning09}. The observed periodic signals are mainly detected from Doppler velocity measurements and can therefore be associated to the transverse displacement of the fine structures \citep{Lin09}.  Extensive reviews on small amplitude oscillations in prominences can be found in \cite{OB02,Engvold04,Wiehr04,Ballester06,banerjee07,Engvold08,Oliver09, Ballester10}, and \citet{Mck10}.

Small amplitude oscillations in quiescent filaments have been interpreted in terms of magnetohydrodynamic (MHD) waves \citep{OB02,Ballester06}, which has allowed to develop the theoretical models (see \citealt{Ballester05,Ballester06}, for recent reviews). Models of filament threads usually represent them as part of a larger magnetic flux tube, whose foot-points are anchored in the solar photosphere \citep{BP89,Rempel99}. Early works studying filament thread oscillations have considered the MHD eigenmodes supported by a filament thread modeled as a Cartesian slab, partially filled with prominence plasma, and embedded in the corona \citep{JNR97,diaz01,diaz03}. These works were later extended by considering a more representative cylindrical geometry by \cite{diaz02}. These authors found  that the fundamental transverse fast mode is  always confined in the dense part of the flux tube, hence, for an oscillating cylindrical filament thread, it should be difficult to induce oscillations in adjacent threads, unless they are very close. Groups of multithread structures and their collective oscillations have been modeled by \cite{diaz05,DR06} in  Cartesian geometry and by \cite{Soler09noad} in cylindrical geometry. 
 
Time and spatial damping is a recurrently observed characteristic of small amplitude prominence oscillations. Observational evidence for the damping of small amplitude oscillations in prominences can be found in \cite{Landman77,Tsubaki86,Tsubaki88,Wiehr89,Molowny-Horas99,Terradas02}, and more recently in \cite{Linthesis,Berger08,Ning09,Lin09}. These observational studies have allowed to obtain some characteristic spatial and time scales. Reliable values for the damping time have been derived, from different Doppler velocity time series, by \cite{Molowny-Horas99} and \cite{Terradas02}, in prominences, and by \cite{Linthesis} in filaments.  The values thus obtained are usually between 1 and 4 times the corresponding period, and large regions of prominences/filaments display similar damping times. Several theoretical mechanisms have been proposed in order to explain the observed damping.  \citet{Ballai03} estimated, through order of magnitude calculations, that
several isotropic and anisotropic dissipative mechanisms, such as viscosity, magnetic diffusivity, radiative losses and thermal conduction cannot in general explain the observed wave damping. Linear non-adiabatic MHD waves have been studied by \cite{Carbonell04,Terradas01,Terradas05,Soler07a,Soler08}. The overall conclusion from these studies is that thermal mechanisms can only account for the damping of slow waves in an efficient manner, while fast waves remain almost undamped.  Since prominences can be considered as partially ionized plasmas, a possible mechanism to damp fast waves (as well as Alfv\'en waves) could come from ion-neutral collisions \citep{Forteza07,Forteza08}, although the ratio of the damping time to the period does not completely match the observations.  Besides non-ideal mechanisms,  another possibility to attenuate fast waves in thin filament threads comes from resonant wave damping (see e.g.\ \citealt{Goossens10}). This phenomenon is well studied for transverse kink waves in coronal loops (\citealt{GAA06,Goossens08}) and provides a plausible explanation for quickly damped transverse loop oscillations first observed by TRACE (\citealt{Aschwanden99,Nakariakov99}). The time scales of damping produced by these different mechanisms should be compared with those obtained from observations. The theoretical approach of many works studying the damping of prominence oscillations has been to first study a given damping mechanism in simplified uniform and unbounded media thereafter introducing structuring and non-uniformity. This has led to an increasing complexity of theoretical models in such a way that some of them now combine different damping mechanisms.

This paper presents results from recent theoretical studies on small amplitude oscillations in prominences and their damping.  This topic has been the theme of several recent review works, see e.g. \citet{Oliver09} and \citet{Ballester10}, a fact that reflects the liveliness of the subject. Here we extend these works by reporting the last studies and mainly focusing on the theory of damping mechanisms for  oscillations in prominences and its application to prominence seismology, a discipline in a rapidly developing stage, as better observations and more accurate theoretical models  become available. 

\section{Damping of oscillations by thermal mechanisms}
\label{sec:2}

In a seminal paper, \citet{Field1965} studied the thermal instability of a dilute gas in mechanical and thermal equilibrium. Using this approach, the time and spatial damping of magnetohydrodynamic waves in unbounded media \citep{Carbonell04, Carbonell09} and in bounded slabs \citep{Terradas01, Terradas05} with physical properties akin to those of quiescent solar prominences, as well as in slabs having prominence-corona physical properties \citep{Soler07a, Soler09trans} have been studied. Furthermore, the behavior of non-adiabatic waves in an isolated and flowing prominence fine structure \citep{Soler08} has also been  considered. All these investigations have been already reviewed in \citet{Oliver09}, \citet{Ballester10} and \citet{Mck10}. For this reason, in the following we concentrate in a recent investigation about the time damping of linear non-adiabatic MHD waves in a multi-thread prominence when mass flows are also present.

\subsection{Time damping of non-adiabatic magnetoacoustic waves in filament threads with mass flows}
\label{nonadiabatic}

In an attempt to explain the observations about in-phase oscillations of large areas of a filament reported by \citet{Linthesis}, \citet{diaz05} studied the collective oscillations of a set of  inhomogeneous threads in Cartesian geometry. The configuration was made of five unequally separated threads having different widths and densities, while the magnetic field keeps the same strength everywhere. When realistic separations between threads are considered, the system oscillates with the only non-leaky mode, and oscillations are in phase with similar amplitudes and with a frequency smaller than that of the densest thread. Although these results show some agreement with observations, the use of Cartesian geometry favors collective oscillations since the transversal velocity perturbation has a very long tail  which allows for an easy interaction between neighboring threads. However, \citet{Linthesis} also pointed out that, after four periods, the oscillations were strongly damped in time and that simultaneously flowing and oscillatory structures were also present in filament threads. To model this situation, \citet{Soler09noad} have used the T-matrix theory of scattering to study the propagation of non-adiabatic magnetoacoustic waves in an arbitrary system of  cylindrical threads when material flows inside them are present. In this study, the effects of radiative losses, thermal conduction and heating have been taken into account, mass flows along magnetic field lines have been allowed, and  the general case $\beta \neq 0$ has been considered. The equilibrium system is made of two homogeneous and unlimited parallel cylinders, prominence threads, embedded in an also homogeneous and unbounded coronal medium (see Fig.~{\ref{fig:model}). For the study, two different configurations have been chosen. 

\subsubsection{Identical threads}
\label{nonadiabatic_I}

In this case the same typical values of prominence density and temperature have been used for both threads, as well as typical coronal values for the external density and temperature. Furthermore,  the magnetic field strength is the same everywhere. In the absence of flow, four fast kink modes, already found by \citet{Luna2008} when studying the collective oscillations of two coronal loops, are present. 

These modes are named as $S_x, A_x, S_y, A_y$, where $S$ and $A$ denote symmetry or antisymmetry of the  velocity field inside the tubes with respect to the ($y$, $z$) plane and the subscripts refer to the main direction of polarization of motions. In addition to the kink modes, two further fundamental collective wave modes, one symmetric ($S_z$) and one antisymmetric ($A_z$), mainly polarized along the $z$-direction and corresponding to slow modes have been identified.

\begin{figure}
\centering
\includegraphics[width=0.6\textwidth]{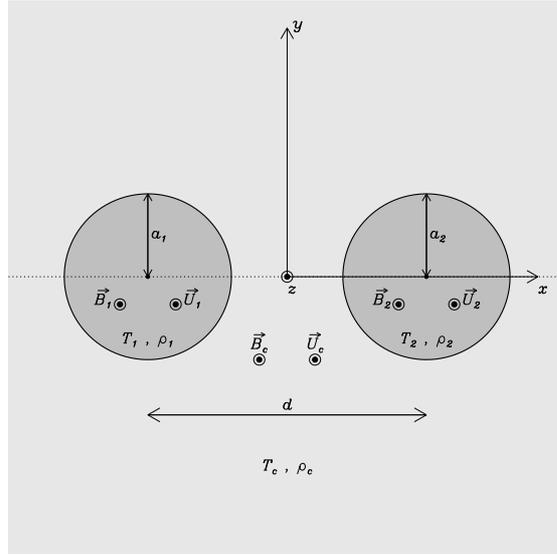}
\caption{Scheme in the $xy$-plane of the model considered in Sect.~\ref{nonadiabatic_I}. 
The $z$- axis is perpendicular to the plane of the figure and points towards the reader. Adapted from \citet{Soler09noad}.}
\label{fig:model}
\end{figure}

Figure~\ref{fig:distkink}a displays the ratio of the real part of the frequency of the four kink solutions to the frequency of the individual kink mode \citep{Soler08}, $\omega_k$,  as a function of the distance between the center of cylinders, $d$. It can be seen that the smaller the distance between centers, the larger the interaction between threads and the separation between frequencies. The frequency of collective kink modes is almost identical to the individual kink frequency for a distance between threads larger than 6 or 7 radii. Therefore, we expect the collective behavior of oscillations to be stronger when the threads are closer. On the opposite, for larger distances, the interaction between threads is much weaker and individual oscillations are expected. Furthermore, Figure~\ref{fig:distkink}b shows the ratio of the damping time to the period,  $\tau_D/P$, of the four kink modes as a function of $d$. This ratio appears to be very large suggesting that  dissipation by non-adiabatic mechanisms is not very efficient to damp the fast kink modes which could be responsible of observed filament threads oscillations.

\begin{figure*}[!t]
\includegraphics[width=0.5\textwidth]{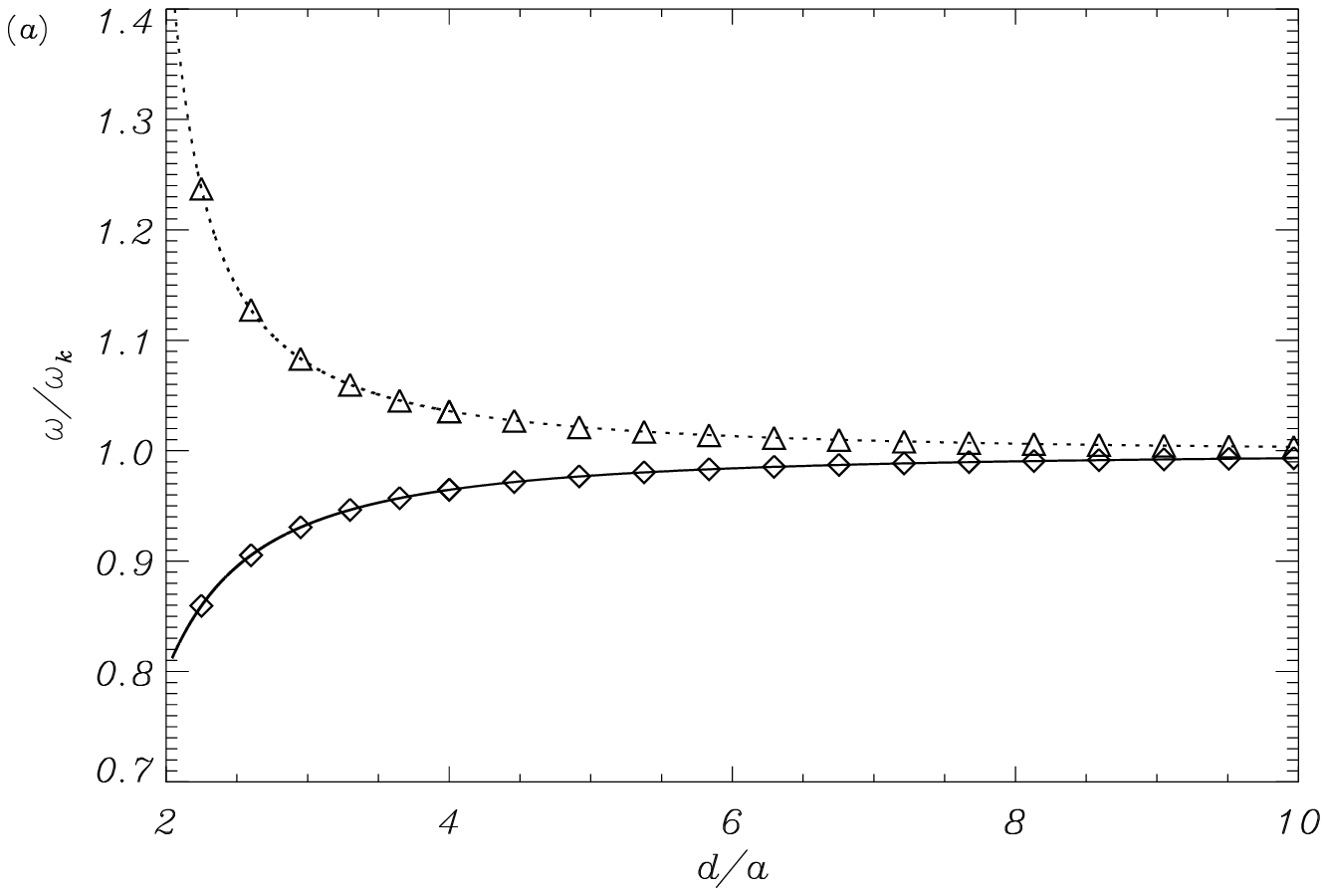}
\includegraphics[width=0.5\textwidth]{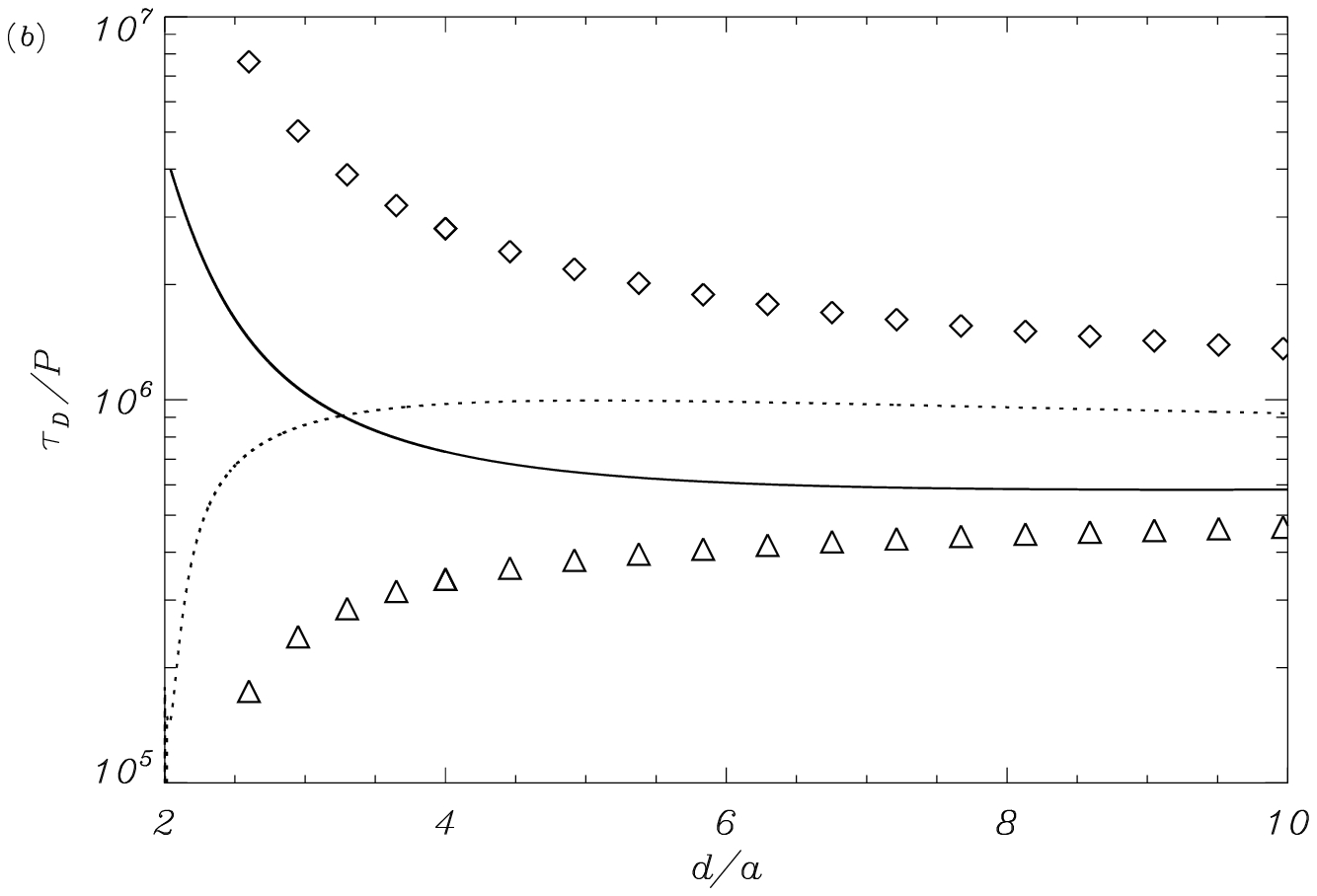}
\caption{$a)$ Ratio of the real part of frequency, $\omega_{\rm R}$, of the $S_x$ (solid line), $A_x$ (dotted line), $S_y$ (triangles), and $A_y$ (diamonds) wave modes to the frequency of the individual kink mode, $\omega_k$, as a function of the distance between centers. $b)$ Ratio of the damping time to the period versus the distance between centers. Linestyles are the same as in panel $a)$. Adapted from \citet{Soler09noad}.}
\label{fig:distkink}
\end{figure*}

In the case of slow modes, Figure~\ref{fig:distslow}a displays the ratio of the real part of the frequency of the $S_z$ and $A_z$ solutions to the frequency of the individual slow mode, $\omega_s$. It can be seen that the frequencies of the $S_z$ and $A_z$ modes are almost identical to the individual slow mode frequency, and the strength of the interaction is almost independent of the distance between cylinders. This is in agreement with the fact that transverse motions (responsible for the interaction between threads) are not significant for slow-like modes in comparison with their longitudinal motions. Therefore, the $S_z$ and $A_z$ modes essentially behave as individual slow modes, contrary to kink modes, which display a more significant collective behavior. Finally, Figure~\ref{fig:distslow}b shows $\tau_D / P$ corresponding to the $S_z$ and $A_z$ solutions versus $d$. Slow modes are efficiently attenuated by non-adiabatic mechanisms, with $\tau_D / P \approx 5$, which is in agreement with previous studies \citep{Soler07a, Soler08} and consistent with observations.

\begin{figure*}[!t]
\includegraphics[width=0.5\textwidth]{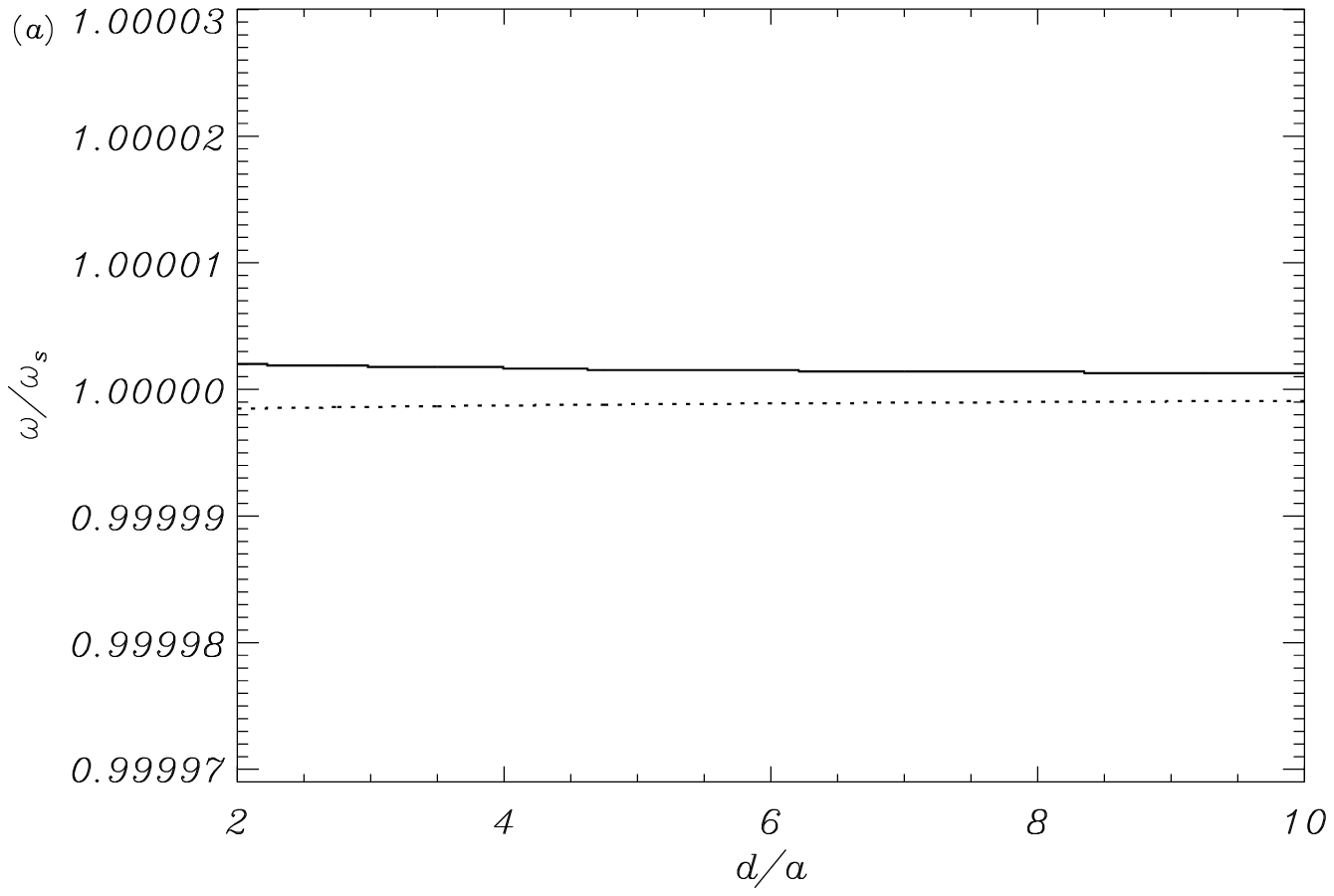}
\includegraphics[width=0.5\textwidth]{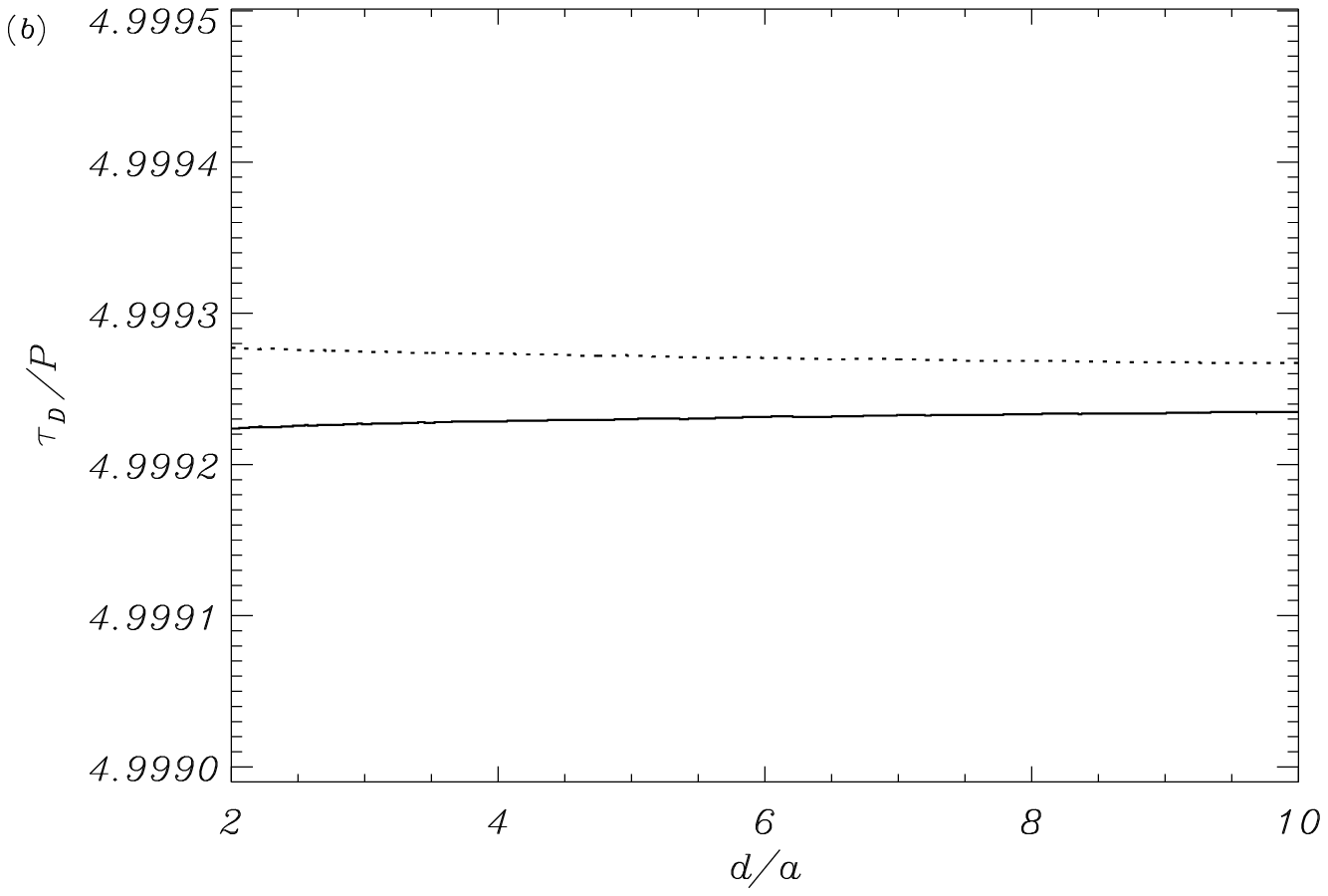}
\caption{$a)$ Ratio of the real part of frequency, $\omega_{\rm R}$, of the $S_z$ (solid line) and $A_z$ (dotted line) wave modes to the frequency of the individual slow mode, $\omega_s$, as a function of the distance between centers. $b)$ Ratio of the damping time to the period versus the distance between centers. Linestyles are the same as in panel $a)$. Adapted from \citet{Soler09noad}.}
 \label{fig:distslow}
\end{figure*}

Next, the effect of flows on the behavior of collective modes has been studied. Arbitrary flows have been assumed in both cylinders while coronal flows have been neglected.  
First of all, we concentrate on transverse modes. We denote the flow in the first cylinder by $U_1$, setting its value to $20$~km~s$^{-1}$, and we study the behavior of the oscillatory frequency when the flow in the second cylinder, denoted by $U_2$, varies (see Fig.~\ref{fig:phase}). Since frequencies are almost degenerate, we follow the notation of \citet{VD2008} and call low-frequency modes the $S_x$ and $A_y$ solutions, while high-frequency modes refer to $A_x$ and $S_y$ solutions. We have restricted ourselves to parallel propagation although the argumentation can be easily extended to anti-parallel waves. In order to understand the asymptotic behavior of frequencies in Figure~\ref{fig:phase}, we define the following Doppler-shifted individual kink frequencies:
\begin{equation}
  \Omega_{k 1} = \omega_k + U_1 k_z, \label{eq:wkleft}
\end{equation}
\begin{equation}
  \Omega_{k 2} = \omega_k + U_2 k_z. \label{eq:wkright}
\end{equation}
Where the $+$ signs arise because of the performed Fourier analysis of the form $\exp(-ik_z z+i \omega t)$. Since $U_1$ is fixed, $\Omega_{k 1}$ is a horizontal line in Figure~\ref{fig:phase}, whereas $\Omega_{k 2}$ is linear with $U_2$.

\begin{figure}[!t]
\centering
\includegraphics[width=0.7\textwidth]{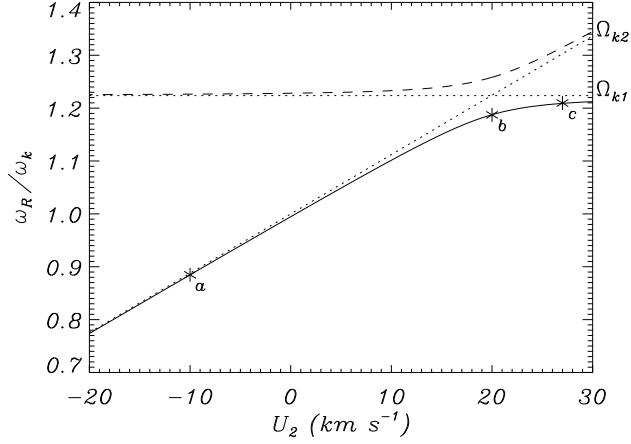}
\caption{Ratio of the real part of the frequency, $\omega_{\rm R}$, to the individual kink frequency, $\omega_k$, as a function of $U_2$ for $U_1=$~20~km~s$^{-1}$. The solid line corresponds to parallel low-frequency modes ($S_x$ and $A_y$) while the dashed line corresponds to parallel high-frequency solutions ($A_x$ and $S_y$). Dotted lines correspond to the Doppler-shifted individual kink frequencies of the threads,  $\Omega_{k 1}$ and $\Omega_{k 2}$. The small letters next to the solid line refer to particular situations studied in the text.  Adapted from \citet{Soler09noad}.}
\label{fig:phase}
\end{figure}

Three interesting situations are worth to study and are denoted as $a$, $b$, and $c$ in Figure~\ref{fig:phase}.

 $(a)$  When $U_2 = -10$~km~s$^{-1}$, which corresponds to a situation with counter-streaming flows, the solutions do not interact with each other and low-frequency (high-frequency) solutions are related to individual oscillations of the second (first) thread. For an external observer this situation would correspond to an individual thread oscillation.

 $(b)$ When $U_2 = 20$~km~s$^{-1}$, both flow velocities and their directions are equal in both threads. In such a situation, there is a coupling between low- and high-frequency modes, and an avoided crossing of the solid and dashed lines is seen in Figure~\ref{fig:phase} and collective oscillations appear.

$(c)$ When $U_2 = 27$~km~s$^{-1}$. This case is just the opposite to $(a)$. Therefore, it corresponds again to an individual thread oscillation.

Considering now slow modes, the behavior of the $S_z$ and $A_z$ solutions can only be considered collective when the flow velocity is the same in both threads because, in such a case, both modes couple. When different flows are considered, the $S_z$ and $A_z$ slow modes behave like individual slow modes. The coupling between slow modes is very sensitive to the flow velocities, and the $S_z$ and $A_z$ solutions quickly decouple if  $U_1$ and $U_2$ slightly differ.

\subsubsection{Non-identical threads}
\label{nonadiabatic_NI}

Consider, now, a system of two nonidentical threads and focus first on kink modes. From above, we expect  collective kink motions to occur when the Doppler-shifted individual kink frequencies of both threads coincide. Following  \citet{Soler09noad}, the relation between flow velocities $U_1$ and $U_2$ for which the coupling takes place is,

\begin{equation}
 U_1 - U_2  \approx \pm \sqrt{2} \left( v_{{\rm A} 2} - v_{{\rm A} 1} \right), \label{eq:velrelation}
\end{equation}
where the $+$ sign is for parallel waves and the $-$ sign is for anti-parallel propagation. A similar analysis can be performed for slow modes to obtain,
\begin{equation}
   U_1 - U_2  \approx \pm \left( c_{s 2} - c_{s 1} \right). \label{eq:velrelationslow}
\end{equation}
which points out that, in general,  the coupling between slow modes occurs at different flow velocities than the coupling between kink modes. Therefore,  the simultaneous existence of collective slow and kink solutions in systems of non-identical threads is difficult.

Then,  we could conclude that the relation between the individual Alfv\'en (sound) speed of threads is determinant for the collective or individual behavior of kink (slow) modes. In the absence of flows and when the Alfv\'en speeds of threads are similar, kink modes are of collective type. On the contrary, when Alfv\'en speeds differ each thread oscillates independently. The same happens for slow modes, but replacing the Alfv\'en speeds by the sound speeds of threads. 

In summary, when flows are present in the equilibrium, collective motions can be found even in systems of non-identical threads by considering appropriate flow velocities. These velocities are within the observed values if threads with not too different temperatures and densities are assumed. However, since the flow velocities required for collective oscillations must take very particular values, such a special situation may rarely occur in real prominences. This conclusion has important repercussions for future prominence seismological applications, in the sense that if collective oscillations are observed in large areas of a prominence \citep{Linthesis}, threads in such regions should possess very similar temperatures, densities, and magnetic field strengths.

This study has also confirmed that collective slow modes are efficiently damped by thermal mechanisms, with damping ratios similar to those reported in observations, $\tau_D / P \approx 5$, while collective fast waves are poorly damped. This is a key point since if the observed efficiently damped oscillations are transverse, this could suggest that other mechanisms could be at work (see Sects.~$\ref{DIN}$ and $\ref{arregui-alfven}$) while if the observed oscillations correspond to slow modes, then, thermal mechanisms could account for the observed damping. Therefore, it becomes crucial to be able to discriminate between the transversal or longitudinal character of the oscillations. However, we should also be  aware that because of the inhomogeneous nature of  filament threads coupling between modes could make a proper mode identification difficult.

\section{Damping of oscillations by ion-neutral collisions}
\label{DIN}

Since the temperature of prominences is typically of the order of 10$^4$ K, the prominence plasma is only partially ionized. The exact ionization degree of prominences is unknown and the reported ratio of electron density to neutral hydrogen density \citep[see e.g.][]{patsovial02} covers about two orders of magnitude (0.1-10). Partial ionization brings the presence of neutrals and electrons in addition to ions, thus collisions between the different species are possible. Because of the occurrence of collisions between electrons with neutral atoms and ions, and more importantly between ions and neutrals, Joule dissipation is enhanced, when compared to the fully ionized case.

The main effects of partial ionization on the properties of MHD waves are manifested through a generalized Ohm's law, which adds some extra terms in the resistive magnetic induction equation, in comparison to the fully ionized case.  The induction equation can be cast as \citep{Soler09rapi}

\begin{eqnarray}
  \frac{\pd {\mathit {\bf B}}_1}{\pd t} &=& \nabla \times \left( {\mathit {\bf v}}_1 \times {\mathit {\bf B}}_0\right) - \nabla \times \left( \eta \nabla \times {\mathit {\bf B}}_1 \right) +  \nabla \times \left\{  \eta_{\rm A} \left[ \left( \nabla \times {\mathit {\bf B}}_1  \right) \times {\mathit {\bf B}}_0 \right] \times {\mathit {\bf B}}_0  \right\}\nonumber \\
 &&-  \nabla \times \left[  \eta_{\rm H} \left( \nabla \times {\mathit {\bf B}}_1  \right) \times {\mathit {\bf B}}_0 \right] , \label{eq:induction}
\end{eqnarray}

\noindent
with ${\mathit {\bf B}}_1$, and ${\mathit {\bf v}}_1$   the perturbed  magnetic field and velocity, respectively. 
Quantities $\eta$, $\eta_{\rm A}$, and $\etah$ in Equation~(\ref{eq:induction}) are the coefficients of ohmic, ambipolar, and Hall's magnetic diffusion, and govern the collisions between the different plasma species. Ohmic diffusion is mainly due to electron-ion collisions, ambipolar diffusion is mostly caused by ion-neutral collisions, and Hall's effect is enhanced by ion-neutral collisions since they tend to decouple ions from the magnetic field while electrons remain able to drift with the magnetic field \citep{Pandey08}. The ambipolar diffusivity can be  expressed in terms of the 
Cowling's coefficient, $\etac$, as 

\begin{equation}
 \eta_A=\frac{\etac-\eta}{\|{\bf B}_0\|^2}.
\end{equation}

\noindent
The quantities  $\eta$ and $\etac$ correspond to the magnetic diffusivities longitudinal and perpendicular to magnetic field lines, respectively. For a fully ionized plasma, $\etac=\eta$, there is no ambipolar diffusion, and magnetic diffusion is isotropic. Due to the presence of neutrals, $\etac \gg \eta$, meaning that perpendicular magnetic diffusion is much more efficient than longitudinal magnetic diffusion in a partially ionized plasma.  It is important to note that $\etac\gg\eta$ even for a small relative density of neutrals.

\subsection{Partial ionization effects in a homogeneous an unbounded prominence medium}

\subsubsection{Time damping of magnetohydrodynamic waves}

Several studies have considered the damping of MHD waves in the solar atmosphere in partially ionized plasmas  \citep{depontieu01,James03,Khodachenko04,Leake05}.
In the context of solar prominences \cite{Forteza07} derived the full set of MHD equations for a partially ionized, one-fluid hydrogen plasma and applied them to the study of the time damping of linear, adiabatic fast and slow magneto-acoustic waves in an unbounded prominence medium.  A partially ionized plasma can be represented as a single-fluid in the strong coupling approximation, which is valid when the ion density in the plasma is small and the collision time between neutrals and ions is short compared with other timescales in the problem. Using this approximation we can describe the very low frequency and large-scale fluid-like behaviour of plasmas \citep {Goossens03}.
The study by \citet{Forteza07} was later extended to the non-adiabatic case, including thermal conduction by neutrals and electrons and radiative losses \citep{Forteza08}. The most important results obtained by \cite{Forteza07} have been summarized recently by \cite{Oliver09} and \citet{Ballester10}. \cite{Forteza07} consider a uniform and unbounded prominence plasma and find that ion-neutral collisions are more important for fast waves, for which the ratio of the damping time to the period is in the range 1 to 10$^5$, than for slow waves, for which values in between 10$^4$ and 10$^8$ are obtained. Fast waves are efficiently damped for moderate values of the ionization fraction, while in a nearly fully ionized plasma, the small amount of neutrals is insufficient to damp the perturbations.

In the above studies, a hydrogen plasma has been considered, however, $90\%$ of the prominence chemical composition is hydrogen while the remaining $10\%$ is helium. Therefore, it is of great interest to know the effect of the presence of helium on the behavior of magnetohydrodynamic waves in a partially ionized plasma with prominence physical properties. This study has been done by \citet{Soler09helium} in a medium like that considered in  \citet{Forteza08}, but composed of hydrogen and helium. The species present in the medium are electrons, protons, neutral hydrogen, neutral helium (HeI), and singly ionized helium (HeII), while the presence of He III is negligible \citep{GL09}. Under such conditions the basic MHD equations for a non-adiabatic, partially ionized, single-fluid plasma have been generalized.

The hydrogen ionization degree is characterized by $\mutilde_{\rm H}$ which varies between $0.5$, for fully ionized hydrogen, and $1$ for fully neutral hydrogen. The helium ionization degree is characterized by $\delta_{\rm He} = \frac{\xi_{{\rm HeII}}}{\xi_{{\rm HeI}}}$, where $\xi_{{\rm HeII}}$ and $\xi_{{\rm HeI}}$ denote the relative densities of single ionized and neutral helium, respectively. Figure~\ref{fig:mhdwaves} displays $\tau_D/P$ as a function of $k$ for the Alfv\'en, fast, and slow waves, and the results corresponding to several helium abundances are compared for hydrogen and helium ionization degrees of $\mutilde_{\rm H} = 0.8$ and $\delta_{\rm He}=0.1$, respectively. We can observe that the presence of helium has a minor effect on the results.  In the case of Alfv\'en and fast waves (Figs.~\ref{fig:mhdwaves}a,b), a critical wavenumber, $k_{\rm c}^{\rm a}$, at which the real part of the frequency becomes zero occurs. In a partially ionized plasma, this critical wavenumber, $k_{\rm c}^{\rm a}$, is given by 

\begin{equation}
k_{\rm c}^{\rm a} = \frac{2 \va}{\cos \theta( \eta_C + \eta \tan^2 \theta)},
\end{equation}

\noindent
where $\va$ is the Alfv\'en speed and $\theta$ the angle between the wavevector and the equilibrium magnetic field. The Cowling's diffusivity, $\eta_C$, depends on the fraction of neutrals and the collisional frequencies between electrons, ions and neutrals. When a fully ionized plasma is considered the numerical value of both diffusivities is the same and in a fully ionized ideal plasma this numerical value is taken equal to zero, therefore,  the critical wavenumber goes to infinity. When partially ionized plasmas are considered, wavenumbers greater than the critical value only produce purely damped perturbations. 
Since Cowling's diffusivity is larger in the presence of helium because of additional collisions of neutral and singly ionized helium species, $k_{\rm c}^{\rm a}$ is shifted toward slightly lower values  than when only hydrogen is considered, so the larger $\xi_{{\rm HeI}}$, the smaller $k_{\rm c}^{\rm a}$.  In the case of the slow wave (Fig.~\ref{fig:mhdwaves}c), the maximum and the right hand side minimum of $\tau_D/P$ are also slightly shifted toward lower values of $k$. Previous results from \citet{Carbonell04} and \citet{Forteza08} suggest that thermal conduction is responsible for these maxima and minima of $\tau_D/P$. The additional contribution of neutral helium atoms to thermal conduction produces this displacement of the curve of $\tau_D/P$. In the case of Alfv\'en and fast waves, this effect is not very important. Although \citet{GL09} suggest that for central prominence temperature a realistic ratio between the number densities of $\xi_{{\rm HeII}}$ to $\xi_{{\rm HeI}}$ is $10\%$, in Figure~\ref{fig:mhdwaves}, and for sake of comparison, the results for $\xi_{{\rm HeI}} = 10\%$ and $\delta_{\rm He} = 0.5$ have been also plotted.

\begin{figure*}[!t]
\includegraphics[width=0.5\textwidth]{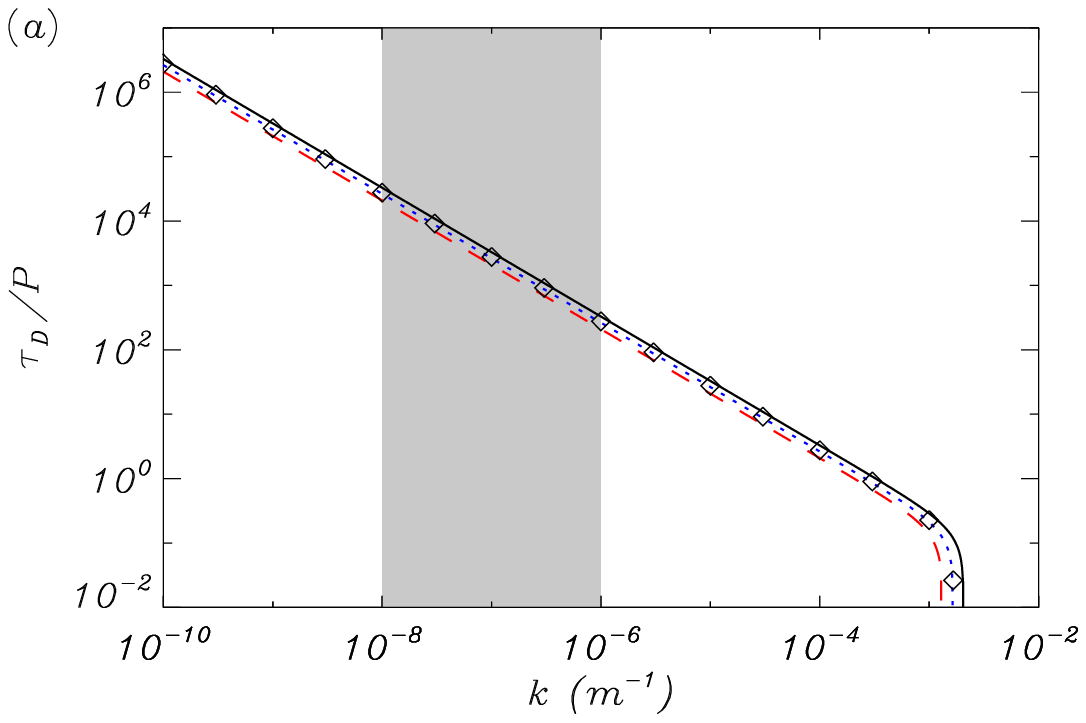}
\includegraphics[width=0.5\textwidth]{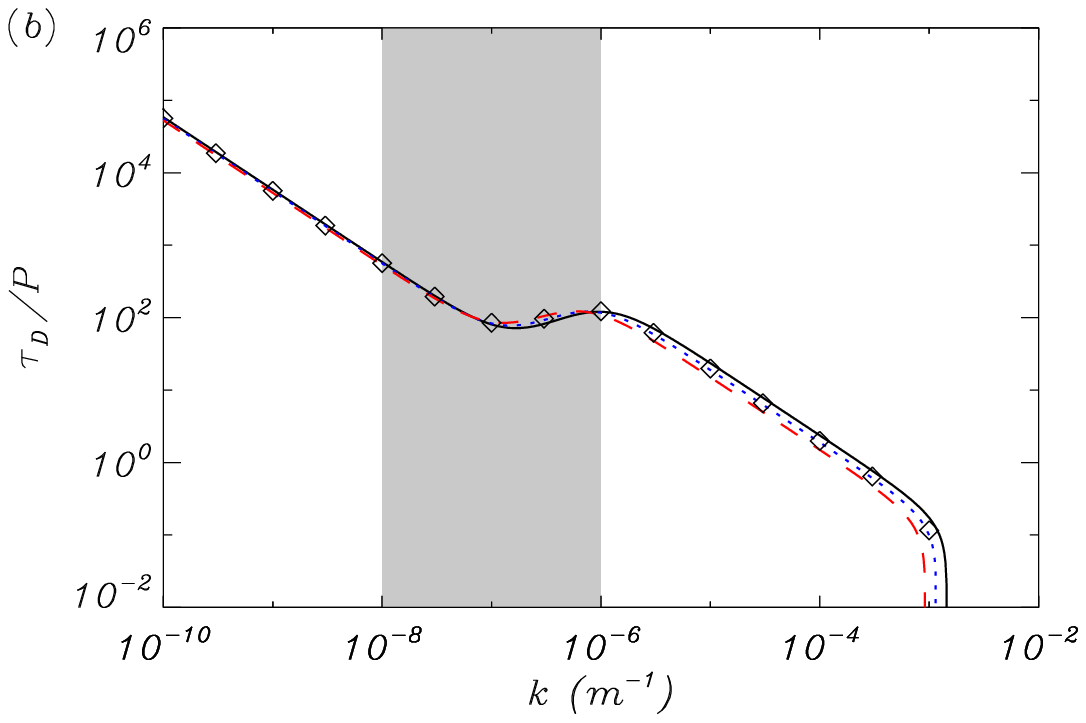}\\
\includegraphics[width=0.5\textwidth]{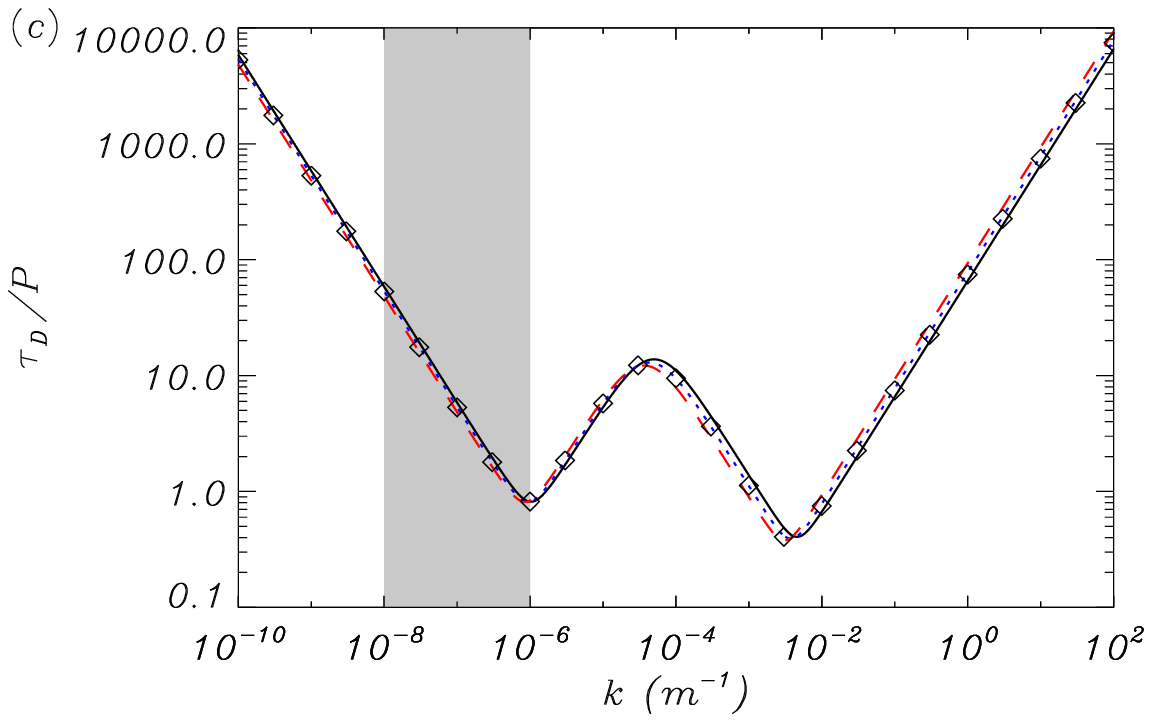}
\includegraphics[width=0.5\textwidth]{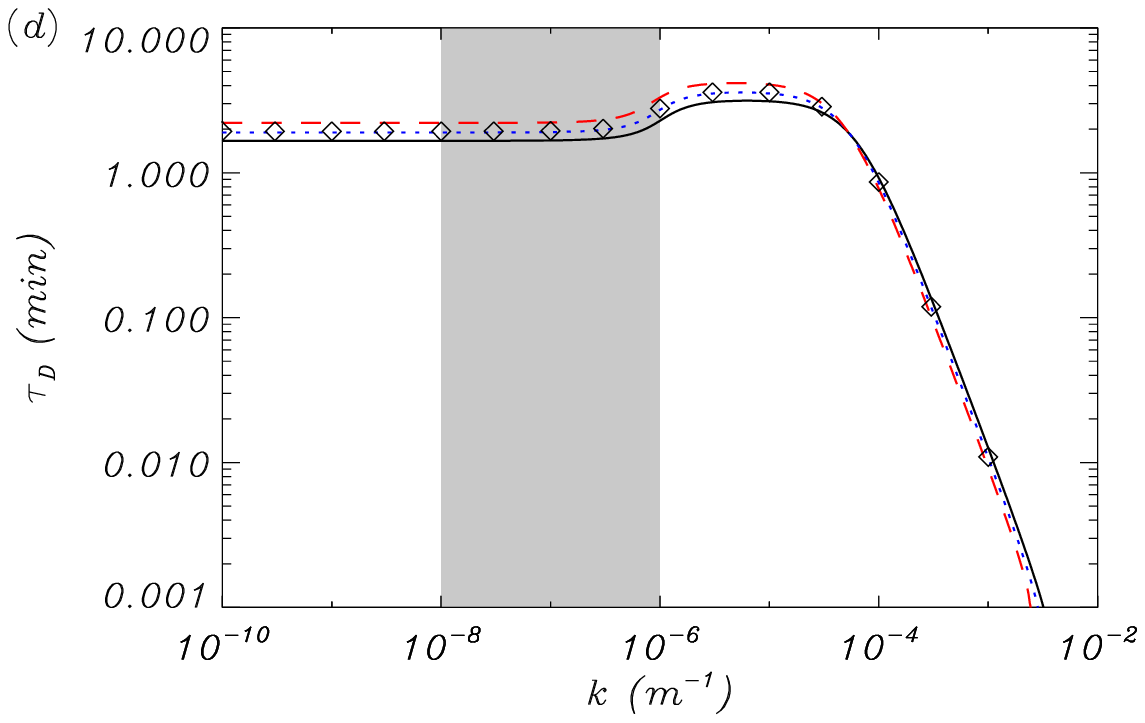}
\caption{($a$)--($c$): Ratio of the damping time to the period, $\tau_D/P$, versus the wavenumber, $k$, corresponding to the Alfv\'en wave,  fast wave, and  slow wave, respectively. 
($d$) Damping time, $\td$, of the thermal wave versus the wavenumber, $k$. The different linestyles represent: $\xi_{{\rm HeI}} = 0\%$ (solid line), $\xi_{{\rm HeI}} = 10\%$ (dotted line), and $\xi_{{\rm HeI}} = 20\%$ (dashed line). In all computations, $\mutilde_{\rm H} = 0.8$, $\delta_{\rm He} = 0.1$, and the angle between the wavevector and the equilibrium magnetic field is $\theta=\pi/4$.  The results for $\xi_{{\rm HeI}} = 10\%$ and $\delta_{\rm He} = 0.5$ are plotted by means of symbols for comparison. The shaded regions correspond to the range of typically observed wavelengths of prominence oscillations. Adapted from \citet{Soler09helium}.}
\label{fig:mhdwaves}
\end{figure*} 

Finally, the thermal mode has been considered. Since it is a purely damped, non-pr\-o\-pa\-ga\-ting disturbance ($\omega_{\rm R} = 0$), we only plot  the damping time, $\tau_D$, as a function of $k$ for $\mutilde_{\rm H} = 0.8$ and $\delta_{\rm He}=0.1$ (Fig.~\ref{fig:mhdwaves}d). We observe that the effect of helium is different in two ranges of $k$. For $k > 10^{-4}$~m$^{-1}$, thermal conduction is the dominant damping mechanism, so the larger the amount of helium, the smaller $\tau_D$ because of the enhanced thermal conduction by neutral helium atoms. On the other hand, radiative losses are more relevant for $k < 10^{-4}$~m$^{-1}$. In this region, the thermal mode damping time grows as the helium abundance increases.  Since these variations in the damping time are very small, we again conclude that the damping time obtained in the absence of helium does not significantly change when helium is taken into account. }

In summary, the study by \citet{Soler09helium} points out that the consideration of neutral or single ionized helium in partially ionized prominence plasmas does not modify the behavior of linear, adiabatic or non-adiabatic MHD waves already found by \citet{Forteza07} and \citet{Forteza08}.

\subsubsection{Spatial damping of magnetohydrodynamic waves}

 \citet{Terradas02} analyzed small
amplitude oscillations in a polar crown prominence and reported the presence of a plane propagating wave as well as a standing
wave.  In the case of the propagating wave, which was interpreted
as a slow MHD wave, the amplitude of the oscillations spatially decreased in a substantial way after a distance of $2-5 \times
10^4$ km from the location where wave motion was being generated.  This distance can be considered as a typical spatial
damping length, $L_\mathrm{d}$, of the oscillations. On the other hand, a typical feature in prominence oscillations is the presence of flows which are observed in $H_\alpha$, UV and EUV lines \citep{Labrosse10}.  In $H_\alpha$ quiescent filaments, the observed velocities range from $5$ to $20$ km s$^{-1}$ \citep{ZEM98, Lin03, Lin07} and, because of physical conditions in filament plasma, they seem to be field-aligned. Recently, observations made with Hinode/SOT by \citet{Okamoto07} reported the
presence of synchronous vertical oscillatory motions in the threads of an active region prominence, together with the
presence of flows along the same threads.  However, in limb prominences different kinds of flows are observed and, for
instance, observations made by \citet{Berger08} with Hinode/SOT have revealed a complex dynamics with vertical downflows
and upflows.  

The spatial damping of magnetohydrodynamic waves in a homogeneous and unbounded fully ionized plasma was studied by \citet{Carbonell06}. 
Recently,  the spatial damping of linear non-adiabatic magnetohydrodynamic waves in a homogenous, unbounded, magnetized and flowing partially ionized plasma has been studied by \citet{Carbonell10}. Since we consider a medium with physical properties akin to those of a 
solar prominence, the density is $\rho_\mathrm{0} = 5 
\times 10^{-11}$  kg m$^{-3}$, the  temperature $T_\mathrm{0} = 8000$ \ K, the 
magnetic field $\vert \bf B_\mathrm{0} \vert$= 10 \ G, and, in general, a field-aligned flow with $v_\mathrm{0} = 10$ km s$^{-1}$ has been considered.

\begin{figure}[!t]
\includegraphics[width=0.5\textwidth]{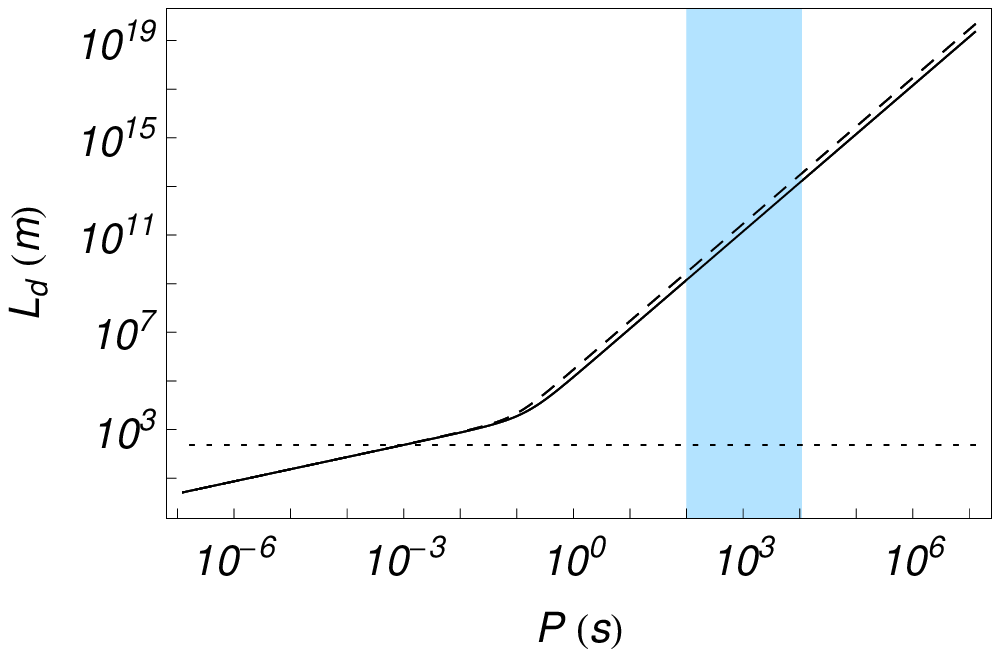}
\includegraphics[width=0.5\textwidth]{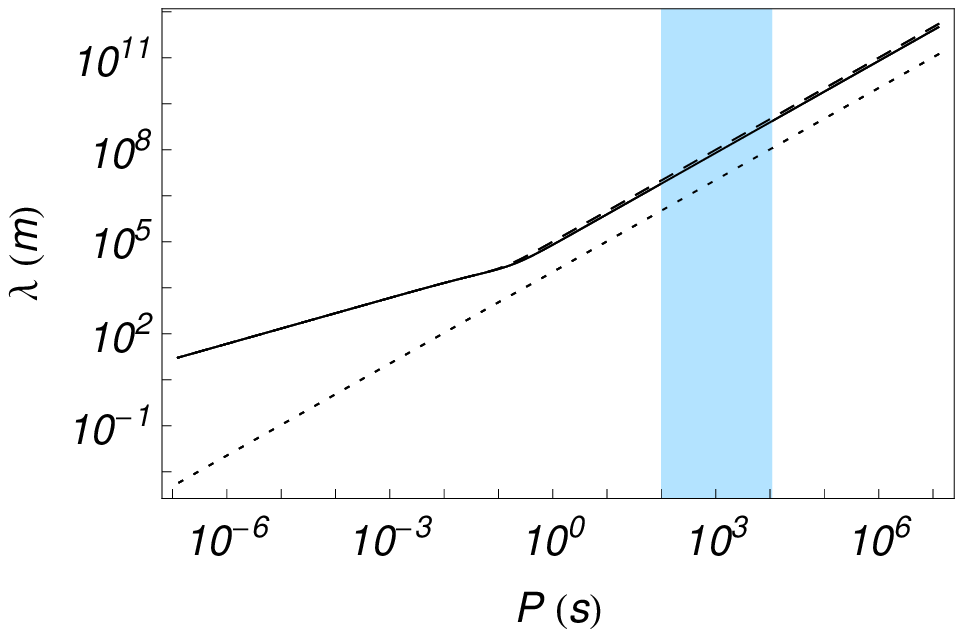}\\
\begin{minipage}{0.5\textwidth}
\includegraphics[width=\textwidth]{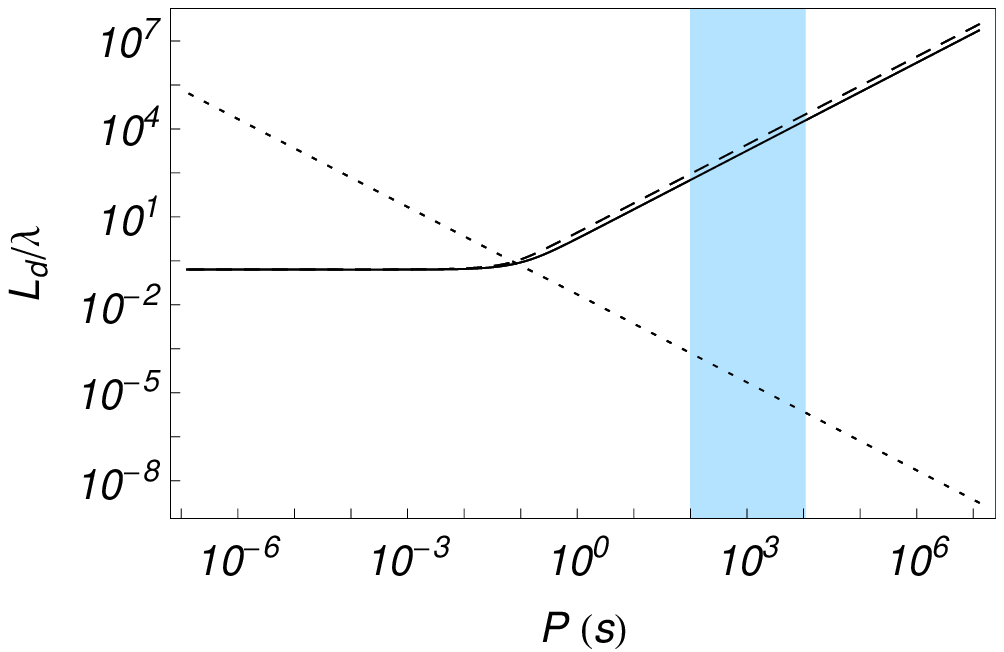}
\end{minipage}
\hspace{0.5cm}
\begin{minipage}{0.45\textwidth}
\caption{Damping length, wavelength, and ratio of the damping length to the wavelength versus period for the three (solid, dashed, dotted) Alfv\'en waves in a partially ionized plasma with an ionization degree $\tilde{\mu}=0.8$ and with a background flow of $10$ \ km s$^{-1}$. In all the panels, the shaded region corresponds to the interval of observed periods in prominence oscillations. \label{f01}}
\end{minipage}

\end{figure}

The dispersion relation for Alfv\'en waves with a background flow is given by:
\begin{eqnarray}
iv_\mathrm{0}(\etac \cos^{2} \theta + \eta  \sin^{2} 
\theta)k^{3} \nonumber \\
+ \left[(v_\mathrm{0}^{2}-v_\mathrm{A}^{2}-i \etac \omega)\cos \theta-i 
\eta \omega \sin \theta \tan \theta\right]k^{2} - 2 \omega 
v_\mathrm{0}k\nonumber{}\\
+ \omega^{2} \sec \theta=0, \label{disp_alf5}
\end{eqnarray}
\noindent
with $\theta$ the angle between the wavevector and the equilibrium magnetic field.
The increase in the degree of the dispersion relation, with respect to the typical dispersion relation for Alfv\'en waves, is produced by the joint presence of flow and resistivities.  In this case, we should obtain three propagating Alfv\'en waves, and Figure~\ref{f01} shows the numerical solution of dispersion relation (\ref{disp_alf5}).   For  the entire interval of periods considered, a strongly damped additional Alfv\'en wave appears, while on the contrary, the other two Alfv\'en waves are very efficiently damped for periods
below $1$ s. However, within the interval of periods typically observed in prominence oscillations these waves are only efficiently attenuated 
 when almost neutral plasmas are considered.

The dispersion relation for thermal and magnetoacoustic waves in presence of a background flow is given by
\begin{eqnarray}
    (\Omega^{2} -k^{2} \Lambda^{2}) (ik^{2} \eta_\mathrm{C} \Omega-\Omega^{2})+k^{2} v_\mathrm{A}^{2}(\Omega^{2} -k_{x}^{2} \Lambda^{2} 
    )+i k^{2} k_{z}^{2}v_\mathrm{A}^{2} \Lambda^{2} \Xi \rho_{0} \Omega= 0,
 \label{disp_mag}
\end{eqnarray}
where $\Lambda^{2}$ is the non-adiabatic sound speed squared \citep{Forteza08, Soler08} and is defined as
\begin{eqnarray}
    \Lambda^{2} = \frac{\frac{T_{0}}{\rho_{0}} A - H +ic_\mathrm{s}^{2} \Omega}{\frac{T_{0}}{p_{0}}A+i \Omega}, \label{nass}
    \end{eqnarray}
where $A$ and $H$, including optically thin radiative losses, thermal conduction by electrons and neutrals, and a constant
heating per unit volume, were defined in \citet{Forteza08}.  When Equation~(\ref{disp_mag}) is expanded, it becomes a seventh
degree polynomial in the wavenumber $k$, whose solutions are three propagating fast waves, two slow waves and two thermal waves.

\begin{figure*}
 \includegraphics[width=0.5\textwidth]{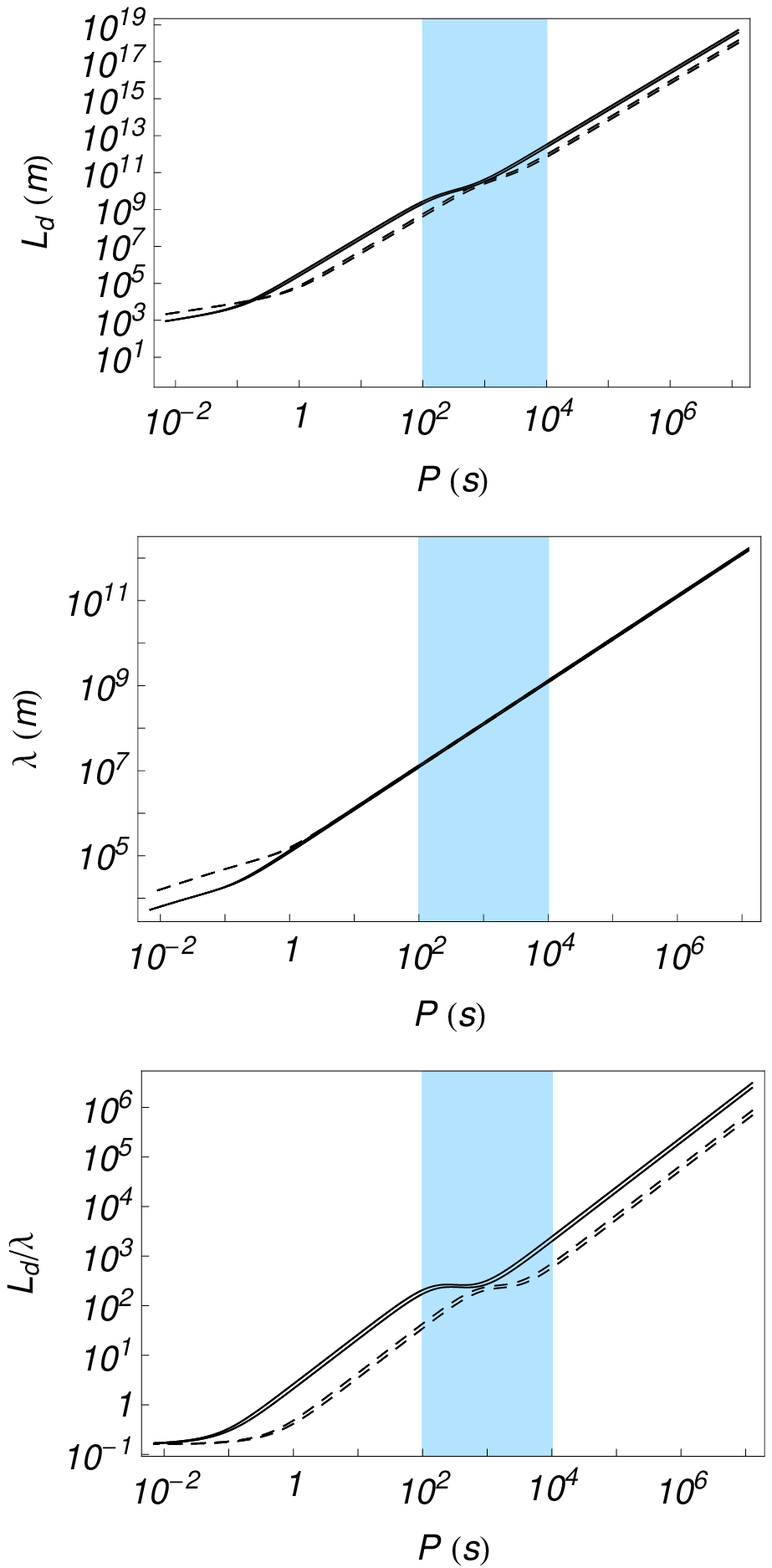}
\includegraphics[width=0.5\textwidth]{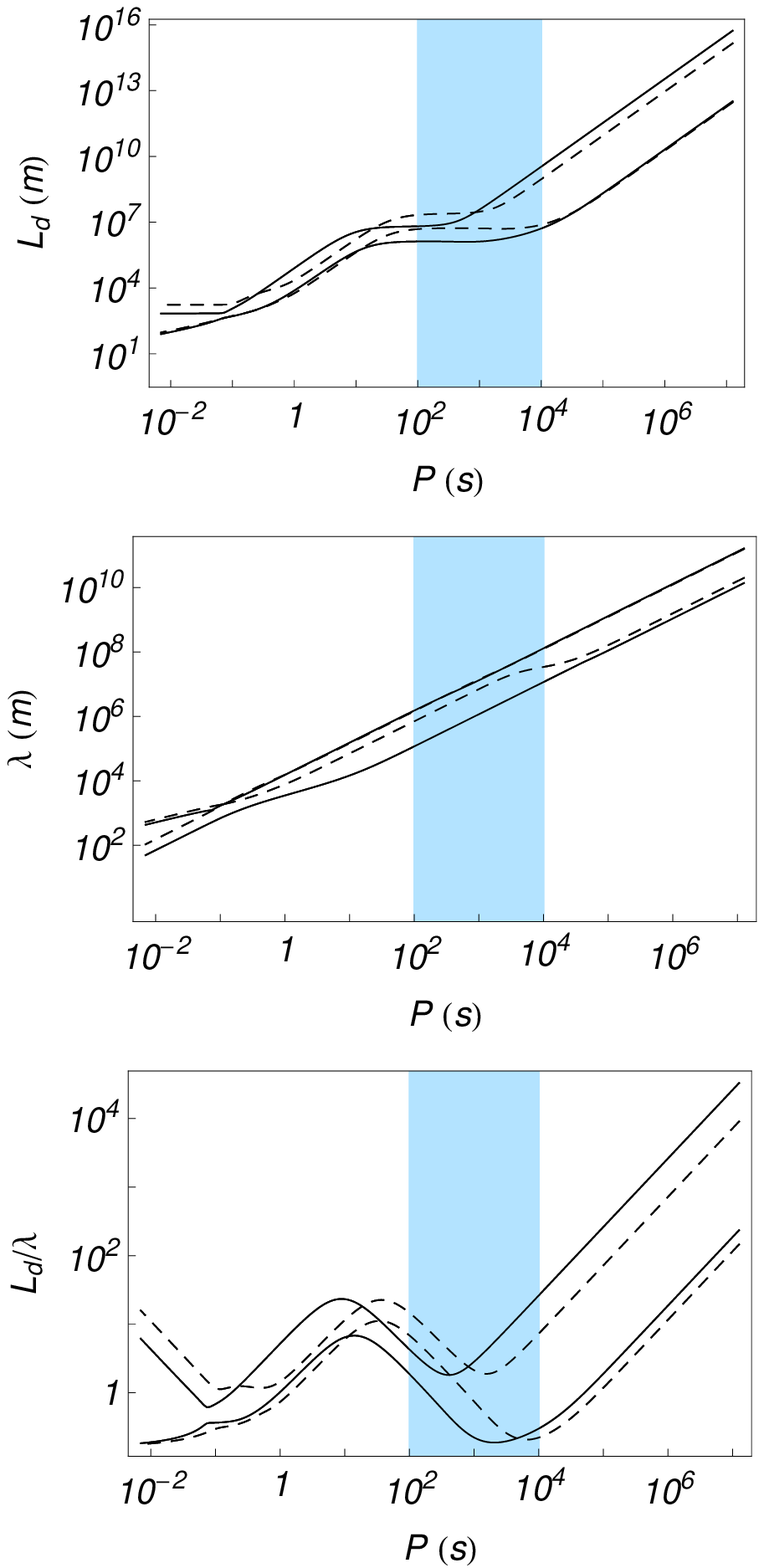} 
		   \caption{Damping length, wavelength, and ratio of the damping length to the wavelength versus period
for the non-adiabatic fast (left panels), slow (right panels) waves in a partially ionized plasma with an ionization degree
 $\tilde{\mu}=0.8$ (solid) and $\tilde{\mu}=0.95$ (dashed). The flow speed is $10$ \ km s$^{-1}$.}   \label{f23} 
 \end{figure*}

Figure~\ref{f23} displays the
behavior of the damping length, wavelength and ratio of damping length versus wavelength for two of the three fast waves  and slow  waves. The curves for the third fast wave, which owes its existence to the joint action of flow and resistivity are similar to those corresponding to the strongly damped Alfv\'en wave in Figure~\ref{f01}.
The most interesting results are those related with the ratio $L_\mathrm{d}/\lambda$.  For fast waves,
this ratio decreases with the period becoming small for periods below $10^{-2}$ s while for one of the slow waves,
 the ratio becomes very small for periods typically observed in prominence oscillations.  When ionization
is decreased, slight changes of the above described behavior occur, the most important being the displacement towards longer
periods of the peak of most efficient damping corresponding to slow waves. 

In summary, Alfv\'en waves in a partially ionized plasma can be spatially damped. When the ionization decreases, the damping length of these waves
also decreases and the efficiency of their spatial damping in the range of periods of interest is improved, although the most
efficient damping is attained for periods below $1$ s.  A new feature is that when a flow is present a
new third Alfv\'en wave, strongly attenuated, appears.  The presence of this wave depends on the joint action of flow and
resistivities,  and it could only be detected by an external observer to the flow.

In the case of non-adiabatic magnetoacoustic waves, when partial ionization is present the behavior of fast, slow and
thermal waves is strongly modified.  In particular,  the damping length of a fast wave in a partially ionized plasma is strongly diminished by neutrals
thermal conduction for periods between $0.01$ and $100$ s, and, at the same time, the radiative plateau present in fully ionized ideal plasmas almost disappears. 
The behavior of slow waves is not so strongly modified as for fast waves, although thermal conduction
by neutrals also diminishes the damping length for periods below $10$ s, and a short radiative plateau still remains for
periods between $10$ and $1000$ s.  Thermal waves are only slightly modified although the effect of partial
ionization is to increase the damping length of these waves, just the opposite to what happens with the other waves.  Next,
when a background flow is included, a new third fast wave appears which, again, is due to the joint action of flow and
resistivities.  Also, in the presence of flow, wavelengths and damping lengths are modified, and since for slow waves sound speed and observed flow speeds are comparable this means that the change in wavelength
and damping length are important, leading to an improvement in the efficiency of the damping.  Moreover, the maximum of efficiency
is displaced towards long periods when the ionization decreases, and for ionization fractions from $0.8$ to $0.95$ it is clearly
located within the range of periods typically observed in prominence oscillations with a value of $L_\mathrm{d}/\lambda$ smaller
than $1$.  This means that for a typical period of $10^{3}$ s, the damping length would be between $10^{2}$ and $10^{3}$ km,
the wavelength around $10^{3}$ km and, as a consequence, in a distance smaller than a wavelength the slow wave would be
strongly attenuated. 

In conclusion, the joint effect of non-adiabaticity, flows and partial ionization allows to damp slow waves in an efficient
way within the interval of periods typically observed in prominences.

\subsection{Partial ionization effects in a cylindrical filament thread model}
\label{robertopicyl}

Recently, \cite{Soler09picyl} have applied the equations derived by \cite{Forteza07} to the study of MHD waves and their time damping in a partially ionized filament thread. The adopted thread model is the classic one-dimensional magnetic flux tube, with prominence conditions, embedded in an unbounded medium with coronal conditions.  A uniform magnetic field, ${\bf B}_0$ pointing along the axis of the tube is considered. As in \cite{Forteza07}, the one fluid approximation for a hydrogen plasma is considered. The internal and external media are characterized by their densities, temperatures, and the relative densities of neutrals, ions, and electrons. The contribution of the latter is neglected. The ionization fraction is thus defined as $\mut=1/(1+\zeta_i)$, with $\zeta_i$ the relative density of ions. For a fully ionized plasma $\mut=0.5$ ($\zeta_i=1$), while for a neutral plasma $\mut=1$ ($\zeta_i=0$).  The external coronal medium is considered as fully ionized, while the ionization fraction in the internal filament plasma is allowed to vary in this range. Note that any value of $\mut$ outside this range is physically meaningless.

In their analysis \cite{Soler09picyl} neglect Hall's term, which is important only for frequencies larger than $\sim$ 10$^4$ Hz, much larger than the observed frequencies of prominence oscillations (for a dimensional analysis that further justifies this approximation see \citealt{Soler09rapi}). After Fourier analyzing the linear MHD wave equations by assuming perturbations of the form $\exp(i \omega t+i m \varphi -i k_z z)$, \cite{Soler09picyl} note that terms with $\etac$ appear accompanied by longitudinal derivatives, while terms with $\eta$ correspond to radial and azimuthal derivatives. By defining $L_{\etac}$ and $L_{\eta}$ as typical length-scales parallel and perpendicular to the magnetic field, the first is associated to the wavelength of perturbations in the longitudinal direction, $\lambda_z\sim 2\pi k_z$, while the second is related to the radius of the structure, $a$. This allows to define the corresponding Reynolds numbers in the parallel and perpendicular directions as $R_{m\parallel}=c_{sf} a/\eta$ and $R_{m\perp}=4\pi^2 c_{sf}/\etac k^2_z a$, where the typical velocity scale has been associated to the sound speed in the filament, $c_{sf}$. The parallel Reynolds number is independent of the wavenumber, while the relative importance of Cowling's diffusion increases with $k_z$. A simple calculation reveals that Cowling diffusion is dominant for the cases of interest. Consider, for instance,  $k_za=1$, $a=100$ km, and $\mut_f=0.8$. Then $R_{m\parallel}=7\cdot 10^6$ and $R_{m\perp}=4\cdot 10^2$.
The wavenumber for which both ohmic and Cowling's diffusion have the same importance can be estimated to be $\lambda\sim[5\cdot 10^3-10^5]$ km, for $a\in[75-375]$ km. In the range of observed wavelengths ($k_za\sim[10^{-3}-10^{-1}]$) both Cowling's and ohmic diffusion could therefore be important. \cite{Soler09picyl} analyze separately the effect(s) of partial ionization in Alfv\'en, fast kink, and slow waves.

\begin{figure*}
 \includegraphics[width=\textwidth]{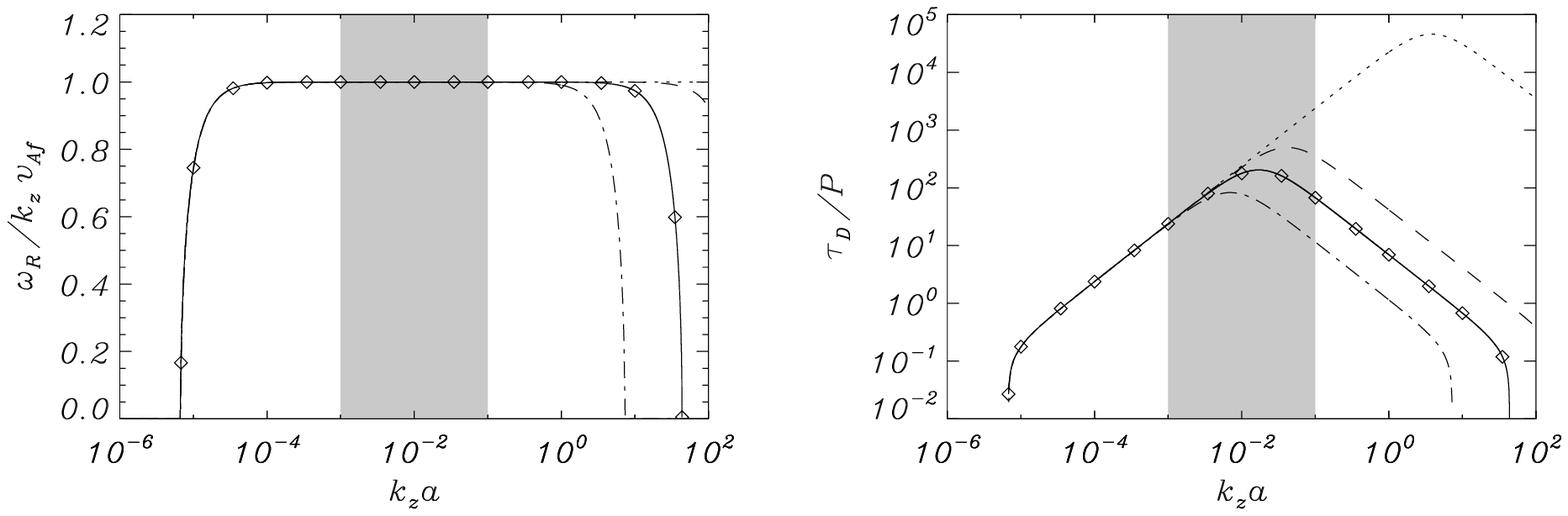}\\
\includegraphics[width=\textwidth]{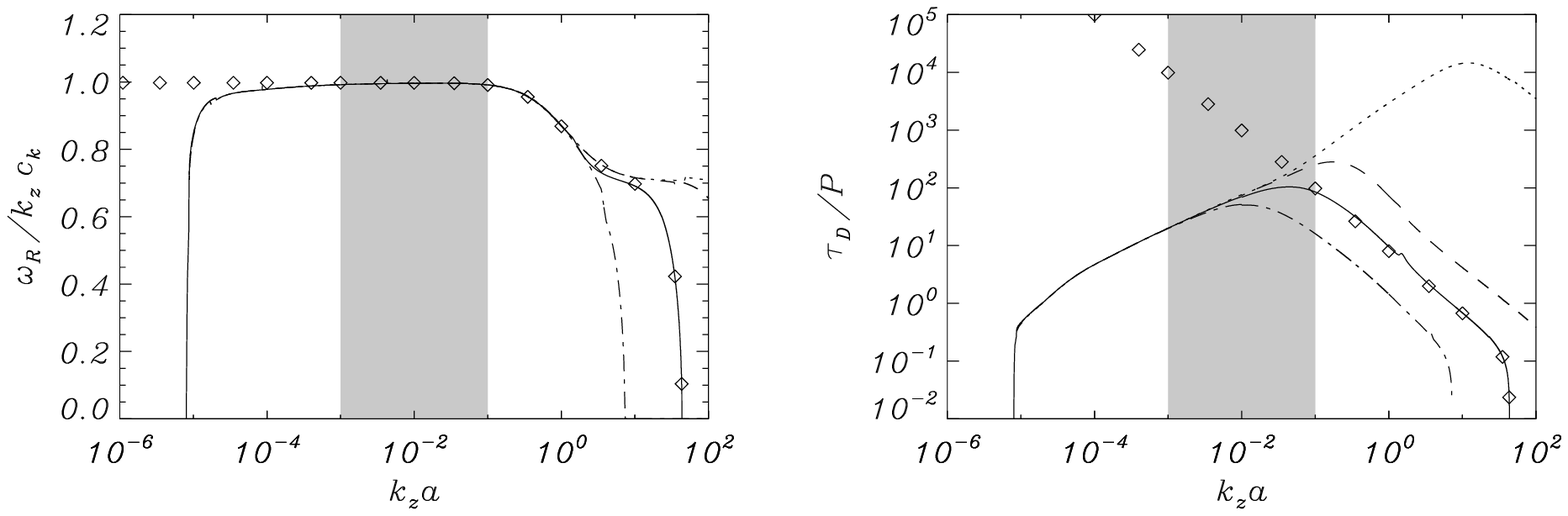}\\
 \includegraphics[width=\textwidth]{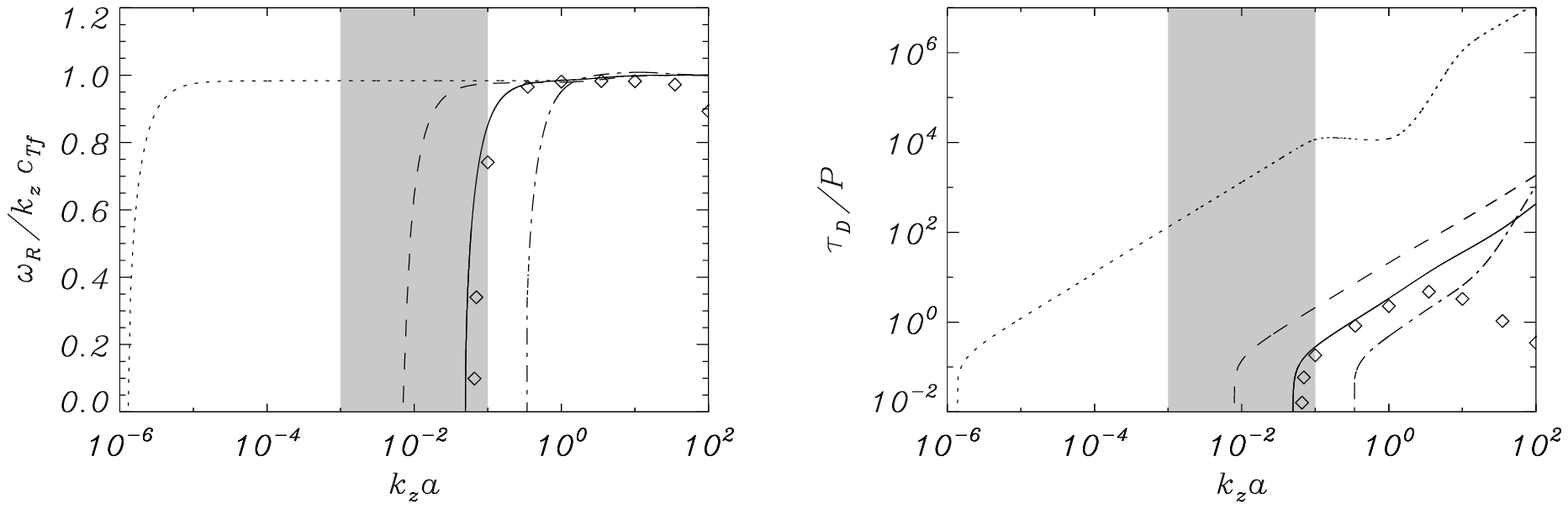}
 \caption{Phase speed (left panels) in units of the Alfv\'en speed, kink speed and internal cusp speed from top to bottom,  and ratio of the damping time to the period (right panels) as a function of $k_z a$ for Alfv\'en (top panels), kink (middle panels) and slow (bottom panels) waves. In all panels different linestyles represent different ionization degrees: $\mut_f=0.5 $ (dotted), $\mut_f=0.6$ (dashed), $\mut_f=0.8$ (solid), and $\mut_f=0.95$ (dash-dotted). Symbols are the approximate solution given by Equation (36) in \citet{Soler09picyl} for $\mut_f=0.8$. The shaded zones correspond to the range of typically observed wavelengths of prominence oscillations. Adapted from \citet{Soler09picyl}. }   \label{figpicyl1} 
 \end{figure*}

\begin{figure}[!t]
\includegraphics[width=0.5\textwidth]{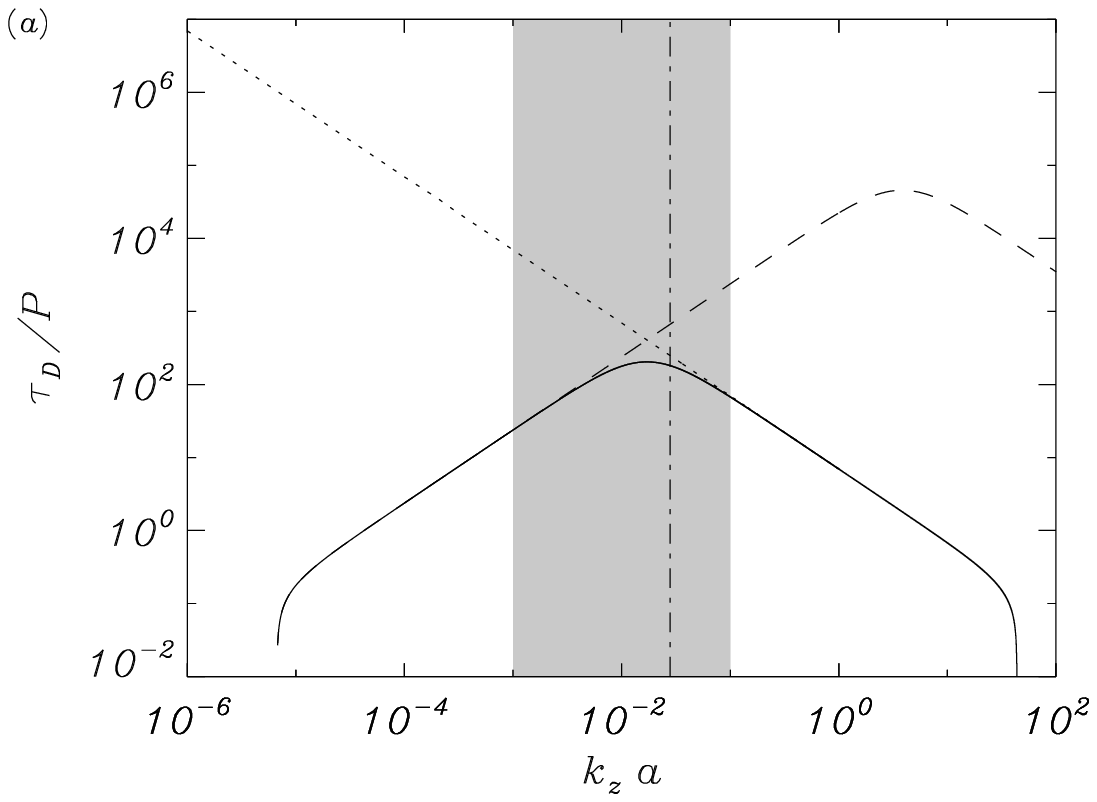}
\includegraphics[width=0.5\textwidth]{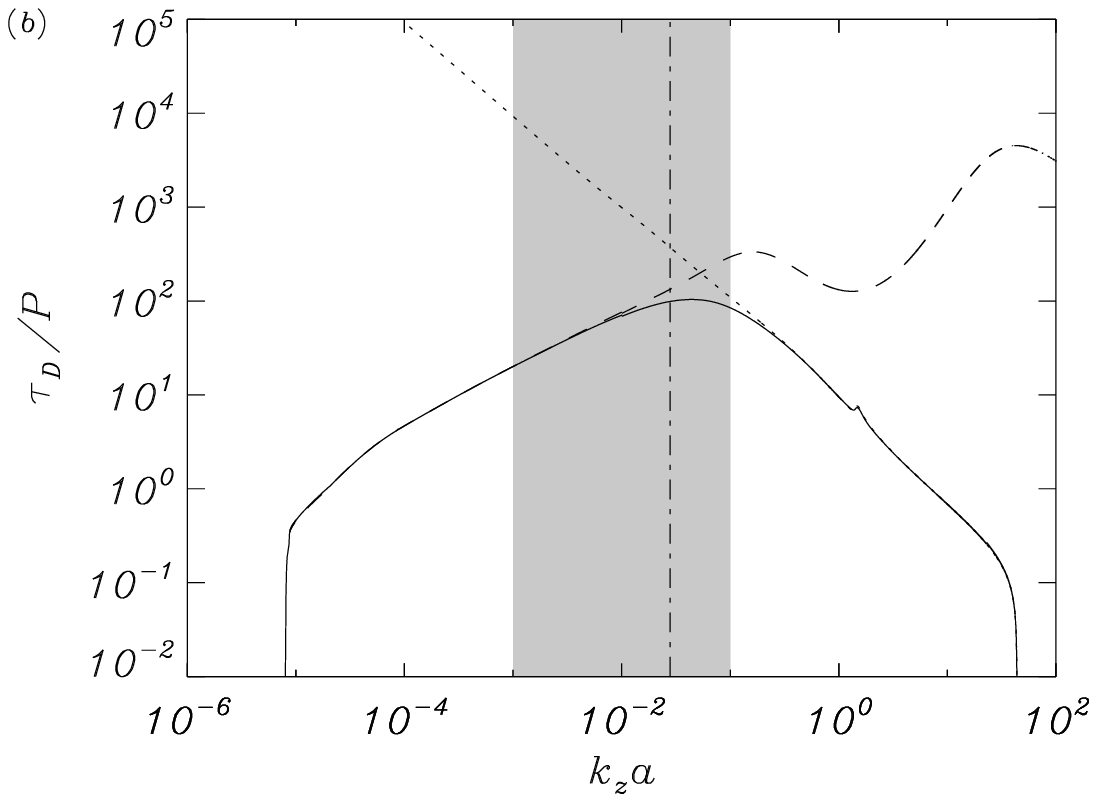}\\
\begin{minipage}{0.5\textwidth}
\includegraphics[width=\textwidth]{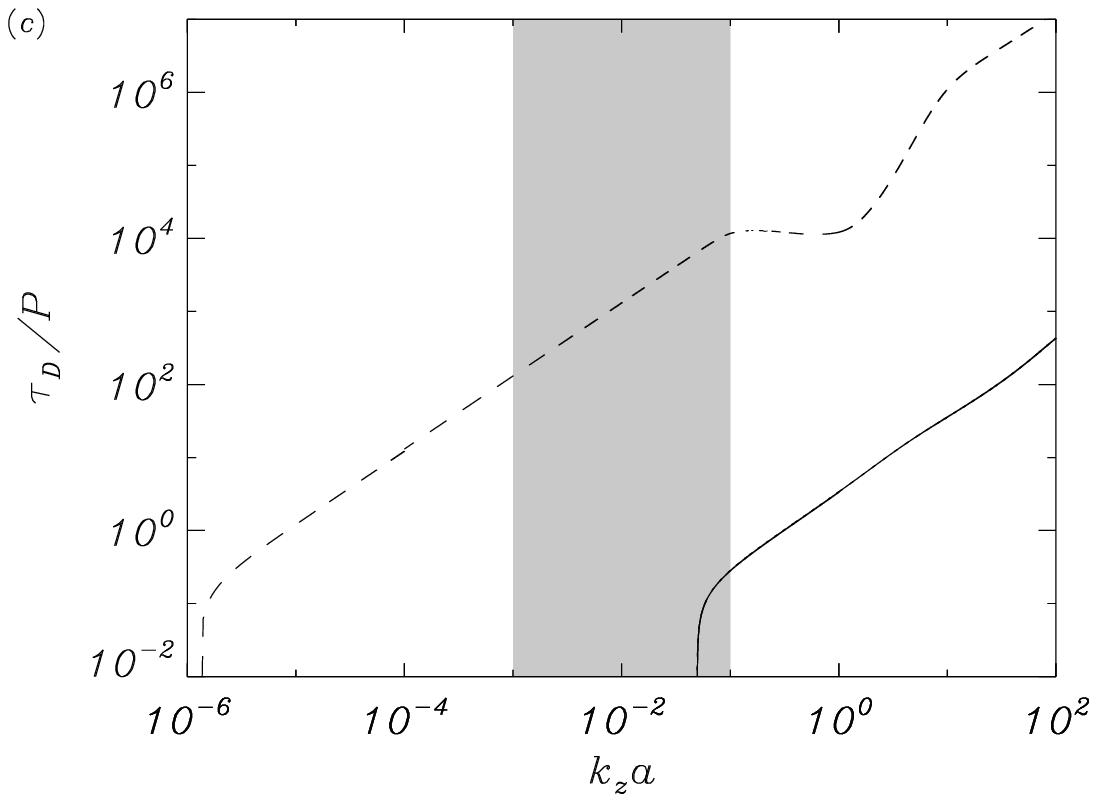}
\end{minipage}
\hspace{0.5cm}
\begin{minipage}{0.45\textwidth}
\caption{Ratio of the damping time to the period as a function of $k_z a$ corresponding to ($a$) the Alfv\'en mode, ($b$), the kink mode, and
($c$) the slow mode, with $\mut_f=0.8$. Solid lines are the complete solution considering all terms in the induction equation. The dotted and dashed lines are the results obtained by neglecting ohmic diffusion ($\eta=0$) or ion-neutral collisions ($\etac=\eta$), respectively. The vertical dot-dashed lines indicate the analytically estimated transitional wavenumber values between both regimes of dominance. Adapted from \citet{Soler09picyl}. \label{figpicyl2}}
\end{minipage}

\end{figure}

Alfv\'en waves are incompressible disturbances with velocity perturbations polarized in the azimuthal direction. In a resistive medium these velocity perturbations are not strictly confined to magnetic surfaces, but have a global nature \citep[see e.g.,][]{FerraroPlumpton61}. \citet{Soler09picyl} show that, by considering solutions with $m=0$ (no azimuthal dependence) and defining a modified Alfv\'en speed squared as $\Gamma^2_A=v^2_A+\imath\omega\etac$, with $v_A$ the ideal Alfv\'en speed, the azimuthal components of the momentum and induction equations can be combined to obtain a Bessel equation of order one for the perturbed magnetic field component in the azimuthal direction. By solving this equation, the phase speed and damping rate of Alfv\'en waves can be studied as a function of the wavelength of perturbations for different ionization degrees and values of the ohmic dissipation. It turns out that Alfv\'en wave propagation is constrained between two critical wavenumber values. These critical wavenumbers are, however, outside the range that corresponds to the observed wavelengths (Fig.~\ref{figpicyl1} top). The small critical wavenumber is found to be insensitive to the ionization fraction, while the large critical wavenumber strongly depends on this parameter. 
The obtained values of the damping time over the period are independent of the ionization degree, for small wavenumber values, while they are affected for large wavenumber values. They are  found to be in between 10 and 10$^2$ times the corresponding period, in the range of typically observed wavelengths. 
By considering solutions to the dispersion relation by neglecting separately one of the two possible damping mechanisms, i.e., partial ionization and ohmic dissipation, \citet{Soler09picyl} observe that the presence of neutrals has an important effect on the damping time for large wavenumbers, while ohmic diffusion dominates for small wavenumbers (Fig.~\ref{figpicyl2}a).

The propagation of transverse kink waves is also found to be constrained by two critical wavenumbers (Fig.~\ref{figpicyl1} middle), a result that can be easily understood by noting that both in the short- and 
long-wavelength regimes the asymptotic approximations for the kink mode frequency directly depend on the Alfv\'en speed. Within the relevant range of observed wavelengths, the phase speed closely corresponds to the ideal counterpart, $c_k=\omega/k_z$, so non-ideal effects are irrelevant to wave propagation for the observed wavelengths. The behavior of the damping rate as a function of wavelength and ionization fraction is seen to closely resemble the result obtained for Alfv\'en waves, with values of $\tau_d/P$ above observed values, in the range of observed wavelengths. Therefore, neither ohmic diffusion nor ion-neutral collisions seem to provide damping times as short as those observed in transverse kink waves in filament threads. Only for an almost neutral plasma, with $\mut_f > 0.95$ and very short wavelengths the obtained damping rates are compatible with the observed time-scales. As for Alfv\'en waves, similar results are obtained for the kink mode when comparing the damping rate by neglecting ohmic diffusion or ion-neutral collisions separately (Fig.~\ref{figpicyl2}b). Ohmic diffusion dominates for small wavenumbers, while ion-neutral collisions are the dominant damping mechanism for large values of $k_za$.

Finally, \citet{Soler09picyl} analyze the propagation properties and the damping by ion-neutral collisions for slow waves. The analysis concentrates on the radially fundamental mode with $m=1$, since the behavior of the slow mode is weakly affected by the value of the azimuthal wavenumber. It turns out that slow wave propagation is constrained by one critical wavenumber, which strongly depends on the ionization fraction,  in such a way that for $k_z<k_{z crit}$ the wave is totally damped. More importantly, for large enough values of $\mut$, the corresponding critical wavelength lies in the range of observed wavelengths of filament oscillations (Fig.~\ref{figpicyl1} bottom). As a consequence, the slow wave might not propagate in realistic thin filament threads. By computing the damping rate, it is found that ion-neutral collisions are a relevant damping mechanism for slow waves, since very short damping times are obtained, especially close to the critical wavenumber. By comparing the  separate contributions of ohmic diffusion and ion-neutral collisions, the slow mode damping is seen to be completely dominated by ion-neutral collisions (Fig.~\ref{figpicyl2}c). Ohmic diffusion is found to be irrelevant, since the presence of the critical wavenumber prevents slow wave propagation for small wavenumbers, where ohmic diffusion would start to dominate.

\section{Resonant damping of filament thread oscillations}\label{resonantdamping}

As discussed in the previous sections, non-adiabatic MHD waves and partial ionization seem to be able to explain the time damping of oscillations in the case of slow waves. The question remains about what physical mechanism(s) could be responsible for the rapid time damping of transverse kink waves in thin filament threads, which seem to be rather insensitive to thermal effects. Partial ionization could be relevant, in view of the results obtained by \cite{Soler09picyl}, but the ratios of the damping time to the period obtained for a cylindrical filament thread are still one or two orders of magnitude larger that those observed. The phenomenon of resonant wave damping in non-uniform media has provided a plausible explanation in the context of quickly damped transverse coronal loop oscillations  \citep{GAA02,Goossens08, Goossens10}. This mechanism relies in the wave coupling and energy transfer between oscillatory modes due to the spatial variation of physical properties that in turn define a non-uniform Alfv\'en and/or cusp speed. Because of the highly inhomogeneous nature of quiescent filament threads at their transverse spatial scales, it is natural to believe that resonant damping must be operative whenever transverse kink waves propagate along those structures. This idea has been put forward by \cite{Arregui08thread}. The analysis by \cite{Arregui08thread} is restricted to the damping of kink oscillations due to the resonant coupling to Alfv\'en waves in a pressureless (zero plasma-$\beta$) plasma. It has been extended to the case in which both the Alfv\'en and the slow resonances are present by \cite{Soler09slowcont}. The main results from these two studies are summarized.

\subsection{Resonant damping in the Alfv\'en continuum}\label{arregui-alfven}

Given the relatively simple structure of filament threads, when compared to the full
prominence/filament system, the magnetic and plasma configuration of an individual 
and isolated thread can be theoretically approximated using a rather simplified model.
\cite{Arregui08thread} make use of a one-dimensional, straight flux tube model in a gravity-free 
environment.  In a system of cylindrical coordinates ($r$, $\varphi$, $z$), with the $z$-axis coinciding with the axis of the tube, the uniform magnetic  field is pointing in the $z$-direction, ${\bf B}=B\ez$. As gas pressure is neglected slow modes are absent and the oscillatory properties of the remaining fast and Alfv\'en MHD waves and their mutual interaction is analyzed. In particular the analysis concentrates on the fundamental kink mode. In such a straight field configuration the zero plasma-$\beta$ approximation implies that the field strength is uniform and that the density profile can be chosen arbitrarily. The non-uniform filament thread is then modeled as a density enhancement with a one-dimensional non-uniform transverse distribution of density, $\rho(r)$, across the structure.  The internal filament plasma, with uniform density,  $\rho_f$, occupies the full length of the tube and is connected to the coronal medium, with uniform density, $\rho_c$, by means of a non-uniform transitional  layer of thickness $l$. If $a$ denotes the radius of the tube, the ratio $l/a$ provides us with a measure of the transverse inhomogeneity length-scale, that can vary in between $l/a=0$ (homogeneous thread) and $l/a=2$ (fully non-uniform thread). The explicit expression for the density profile used by \cite{Arregui08thread} is

\begin{equation}
 \rho_{0}\left(r\right)=\left\{\begin{array}{clc}
 \rho_f,&{\rm if}&r\le a - l/2,   \\
 \rho_{\rm tr}\left(r\right),&{\rm if}&a-l/2<r<a+l/2,\\ 
 \rho_c,&{\rm if}&r\geq  a+l/2, \\
\end{array} \right. \label{rhor}
\end{equation}
with
\begin{equation}
 \rho_{\rm tr}\left(r\right)=\frac{\rho_f}{2}\left\{\left(1+\frac{\rho_c}{\rho_f}\right) - \left( 1-\frac{\rho_c}{\rho_f}\right)\sin \left[\frac{\pi}{l}\left( r-a\right)\right]\right\}.\label{rhotrans}
\end{equation}

\noindent
Typical values for the filament and coronal densities are $\rho_f=5\times10^{-11}~{\rm kg}~{\rm m}^{-3}$ and $\rho_c=2.5\times 10^{-13}~{\rm kg}~{\rm m}^{-3}$, the density contrast between the filament and coronal plasma being $\rho_f/\rho_c=200$.

Observations of transverse oscillations in filament threads can be interpreted in terms of linear kink waves. When considering perturbations of the form $f(r)$exp$(i (\omega t+m\varphi-k_z z))$, with $m$ and $k_z$ the azimuthal and longitudinal wavenumbers and $\omega$  the oscillatory frequency, the fundamental kink mode has $m=1$, and produces the transverse displacement of the tube as it propagates along the density enhancement. This mode is therefore consistent with the detected Doppler velocity measurements and the associated transverse swaying motions of the threads \citep{Lin07,Lin09}.
In the absence of a non-uniform transitional layer, i.e., $l/a=0$, the density is uniform in the internal and external regions and changes discontinuously at the tube radius $a$. Then, a well known dispersion relation is obtained by imposing the continuity of the radial displacement, $\xi_r$, and the total pressure perturbation, $p_T$, at $r=a$. This dispersion relation (Edwin \& Roberts 1983) is 
\begin{equation}
D_m(\omega,k_z)=\frac{\xi_{r,e}}{P'_{T,e}}-\frac{\xi_{r,i}}{P'_{T,i}}=0
\end{equation} 

\noindent
where the indices ``i'' and ``e'' refer to internal and external, respectively, and the prime denotes a derivative with respect to the radial direction. In the commonly used thin tube or long wavelength approximation ($k_za<<1$), the kink mode frequency can be calculated explicitly as

\begin{equation}\label{kinkfrequency}
\omega_k=k_z\sqrt{\frac{\rho_f v^2_{Af}+\rho_c v^2_{Ac}}{\rho_f+\rho_c}},
\end{equation}

\noindent
with $v_{Af,c}=B/\sqrt{\mu\rho_{f,c}}$  the filament and coronal  Alfv\'en velocities. The period of kink oscillations with a wavelength $\lambda=2\pi/k_z$ can be written, in terms of the density contrast, as 

\begin{equation}
P=\frac{\sqrt{2}}{2}\frac{\lambda}{V_{Af}}
\left(\frac{1+c}{c}\right)^{1/2}.\label{period}
\end{equation}

\noindent
Note that the factor containing the density contrast varies between $\sqrt{2}$ and $1$, when $c$ is allowed to vary between a value slightly larger that $1$ (extremely tenuous thread) and 
$c\rightarrow\infty$. This has consequences in the application of the model to perform prominence seismology (see Sect. 6). 

In Equation~(\ref{kinkfrequency}) the eigen-frequency of the fundamental kink mode is in between the internal and external Alfv\'en frequencies, $\omega_{Af}<\omega_k<\omega_{Ac}$, with $\omega_{Af,c}=k_z v_{Af,c}$. This means that when the discontinuous jump in density is replaced by a continuous variation in a non-uniform layer, of thickness $l$, going from $\rho_f$ to $\rho_c$, the fundamental kink mode has its eigen-frequency in the Alfv\'en continuum, and thus couples to an Alfv\'en continuum mode. This results in a transfer of wave energy from the transverse motion of global nature to azimuthal motions of localized nature, and the time damping of the kink mode.
Asymptotic analytical expressions for the damping time, $\tau_d$, can be obtained 
under the assumption that the transverse inhomogeneity length-scale is small ($l/a \ll 1$). This is the so-called thin boundary approximation, which assumes that the thickness of the dissipative layer, namely $\delta_{\rm A}$, coincides with the width of the inhomogeneous transitional layer. This condition is approximately verified for thin layers, i.e., $l/a \ll 1$, which makes the approximation very accurate in such a case. The method was outlined by e.g. \cite{sakurai91,goossens95,TIGO96} and makes use of jump conditions to obtain analytical expressions for the dispersion relation. For the purposes of our description it will suffice to write down these conditions. In a straight and constant magnetic field, we have for the Alfv\'en resonance 

\begin{equation}
[\xi_r]=-\imath\pi\frac{m^2/r^2_A}{\omega^2_A|\partial_r\rho_0|_A}p_T, \mbox{\hspace{1cm}} [p_T]=0, \mbox{\hspace{1cm}} \mbox{\rm at} \mbox{\hspace{1cm}} r=r_A,
\label{jumpalfven}
\end{equation}

\noindent
where $\omega_A$  and $|\partial_r\rho_0|_A$ are the Alfv\'en frequency and the modulus of the radial derivative of the density profile, evaluated at the resonant point. 
These jump conditions allow us to obtain  a dispersion relation of the form

\begin{equation}\label{disperalfven}
D(\omega,k_z)=-\imath\pi\frac{m^2/r^2_A}{\omega^2_A}\frac{\rho_i\rho_e}{|\partial_r\rho_0|_A}.
\end{equation}

\noindent
When the long wavelength and the thin boundary approximations are combined,
the analytical expression for the damping time over period for the kink mode ($m=1$) 
can be written as  (see e.g.\ \citealt{HY88,sakurai91,goossens92,goossens95,RR02})

\begin{equation}
\frac{\displaystyle \tau_{d}}{\displaystyle P} = F \;\; \frac {\displaystyle a}
{\displaystyle l}\;\; \frac{\displaystyle c + 1}{\displaystyle
c - 1}. \label{dampingrate}
\end{equation}

\begin{figure*}[!t]
  \includegraphics[width=0.5\textwidth]{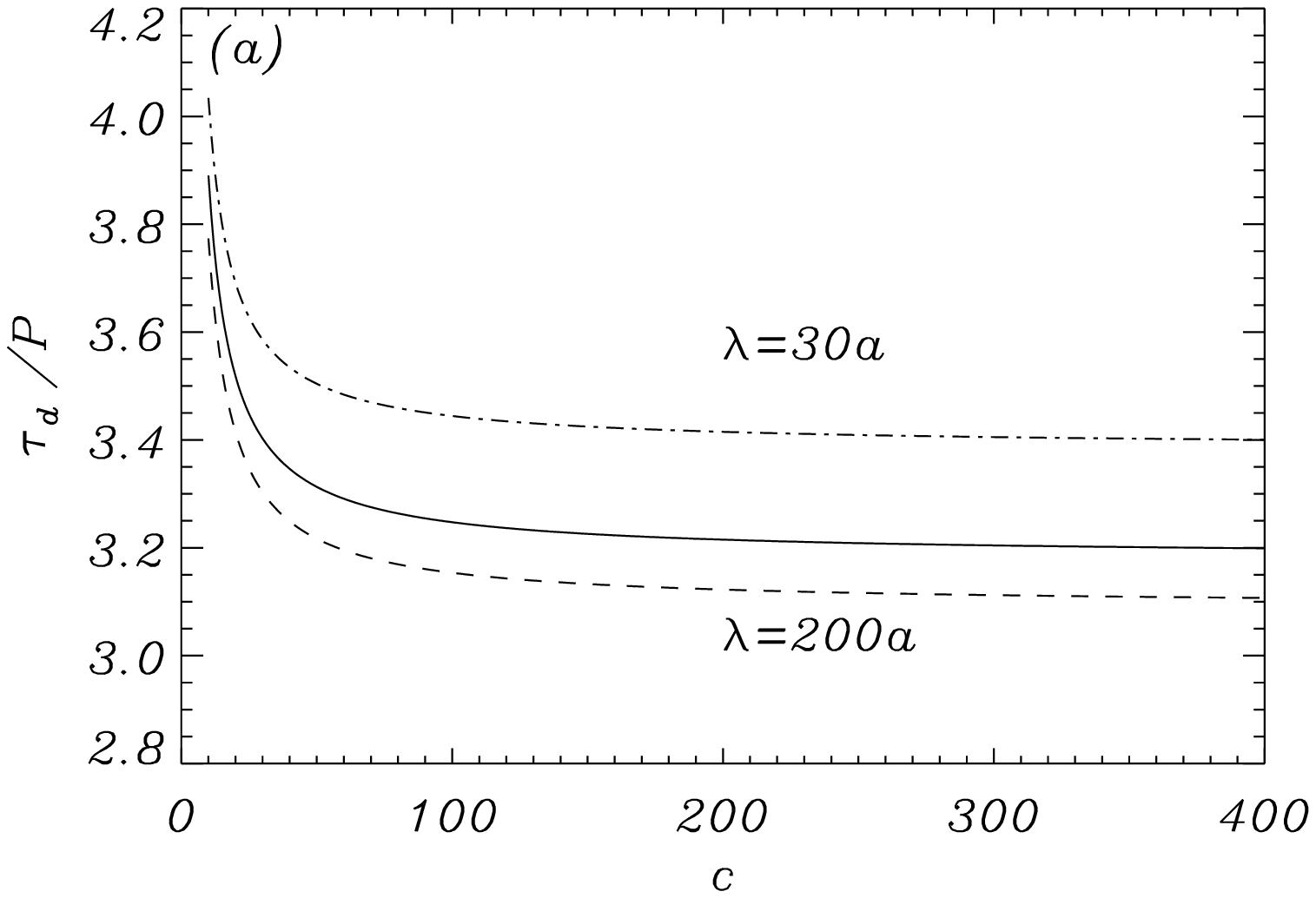}
  \includegraphics[width=0.5\textwidth]{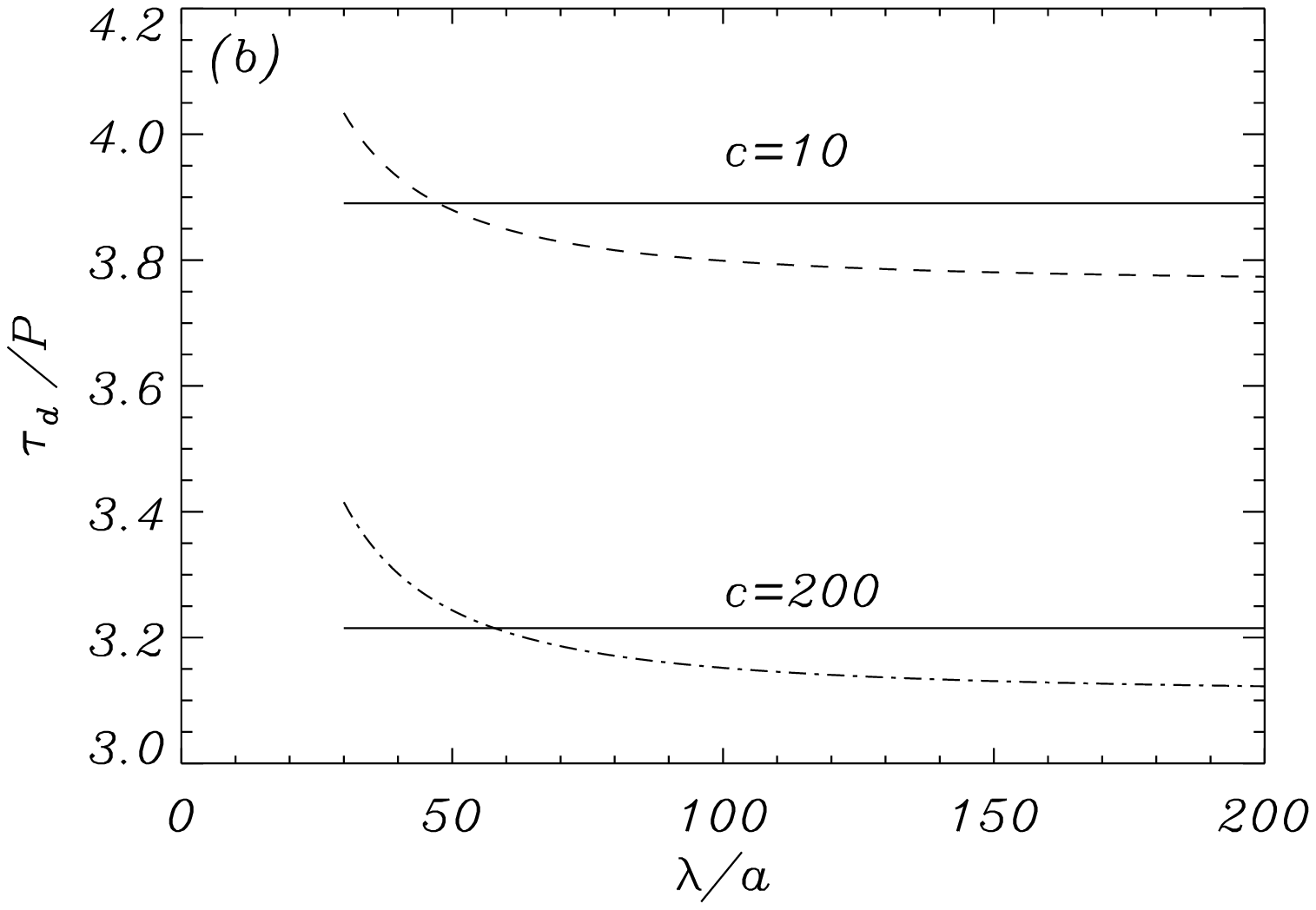}\\
   \includegraphics[width=0.5\textwidth]{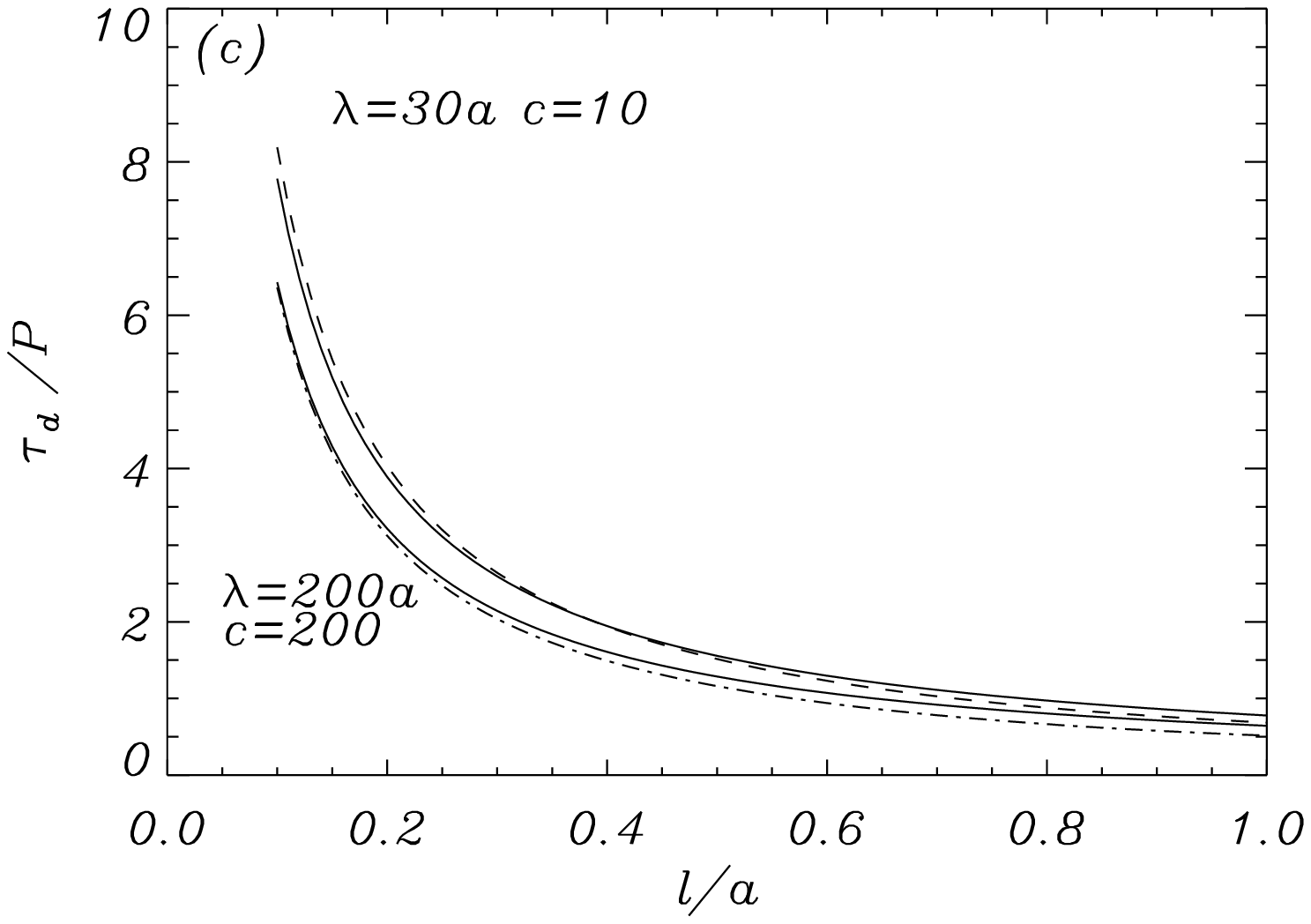}
  \includegraphics[width=0.5\textwidth]{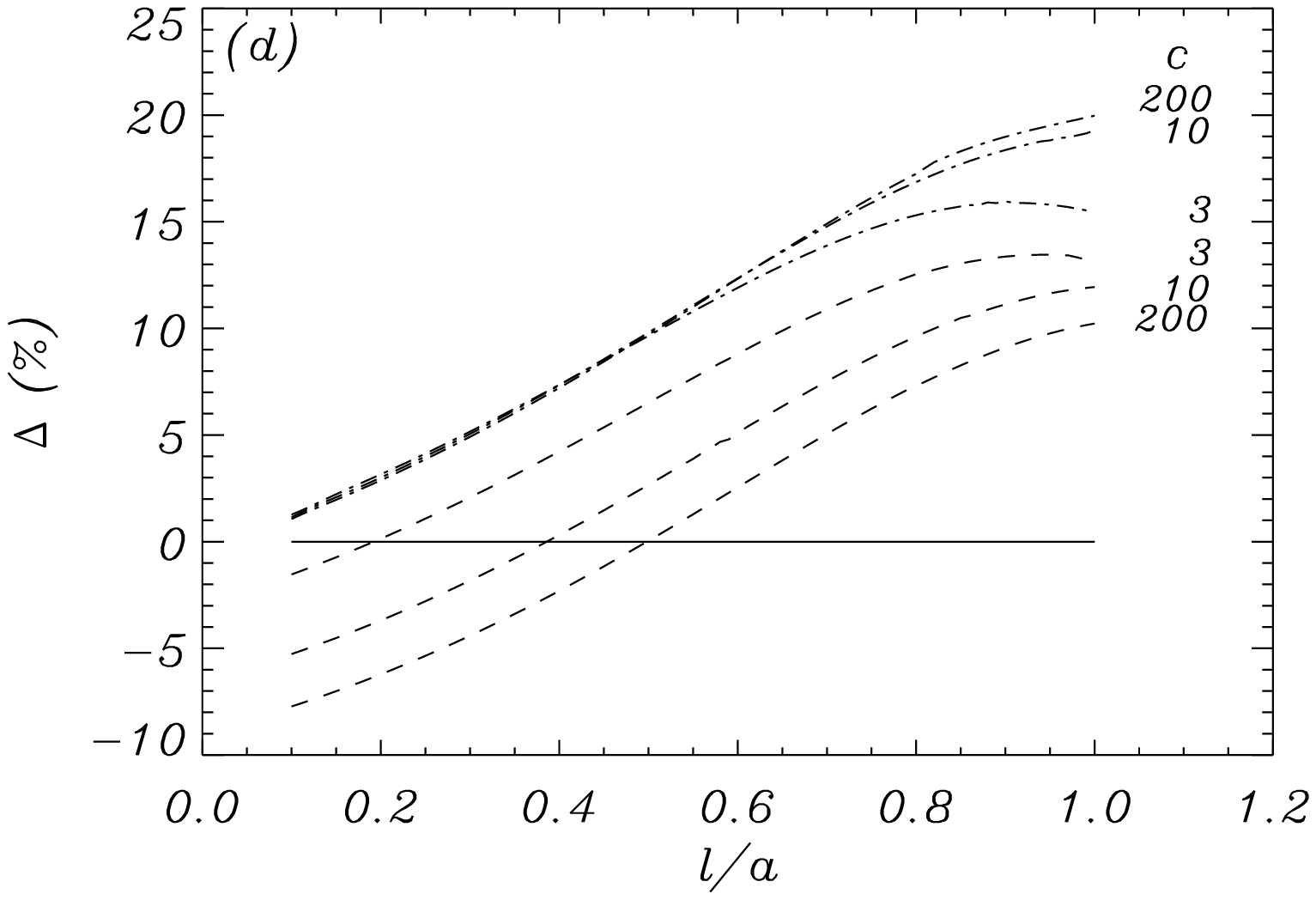}\\
\caption{{\em (a)--(c):} Damping time over period for fast kink waves in filament threads with $a=100$ km.
In all plots solid lines correspond to analytical solutions given by equation~(\ref{dampingrate}), with $F=2/\pi$.
{\em (a):} As a function of density contrast, with $l/a=0.2$ and for two wavelengths. {\em (b)}: 
As a function of wavelength, with $l/a=0.2$, and for two density contrasts. {\em (c)}: As a function of transverse inhomogeneity length-scale, for two combinations of wavelength and density contrast. {\em (d)} Percentage difference, $\Delta$, with respect to analytical formula (\ref{dampingrate}) for different combinations of wavelength, $\lambda=30a$ (dashed lines); $\lambda=200a$ (dash-dotted lines), and density contrast. Adapted from \citet{Arregui08thread}.}
\label{alfven}       
\end{figure*}

\noindent
Here  $F$ is a numerical factor that depends on the particular variation of the density in the non-uniform layer. For a linear variation $F=4/\pi^2$ \citep{HY88,goossens92}; for a sinusoidal variation $F=2/\pi$ \citep{RR02}.  We can already anticipate that, for example, considering $c=200$ as a typical density contrast and $l/a=0.1$ Equation~(\ref{dampingrate}) predicts a damping time  
of $\sim$ 6 times the oscillatory period.

Figure~\ref{alfven} shows analytical estimates computed by \cite{Arregui08thread} using Equation~(\ref{dampingrate}) (solid lines). The damping is affected by the density contrast in the low contrast regime and $\tau_d/P$ rapidly decreases for increasing thread density (Fig.~\ref{alfven}a). Interestingly, it stops being dependent on this parameter in the large contrast regime, typical of filament threads. 
The damping time over period is independent of the wavelength of perturbations (Fig.~\ref{alfven}b), 
but rapidly decreases with increasing  inhomogeneity length-scale (Fig.~\ref{alfven}c).  
These results led \cite{Arregui08thread} to propose that resonant absorption in the Alfv\'en continuum is a very efficient mechanism for the attenuation of fast waves in filament threads, especially because large thread densities and transverse plasma inhomogeneities can be combined together. 
 
The analysis by \cite{Arregui08thread} is completed by computing numerical approximations to the solutions outside the thin tube and thin boundary approximations  used to derive Equations~(\ref{period}) and (\ref{dampingrate}), which may impose limitations to the applicability of the obtained results to filament thread oscillations.  \cite{Arregui08thread} find that analytical and numerical solutions display the same qualitative behavior with density contrast and transverse inhomogeneity length-scale (Figs.~\ref{alfven}a, c). Now the damping time over period slightly depends on the wavelength of perturbations (Fig.~\ref{alfven}b). Equation (\ref{dampingrate}) underestimates/overestimates this magnitude for short/long wavelengths.
The differences are small, of the order of 3\% for $c=10$, and do not vary much with density contrast for long wavelengths ($\lambda=200a$), but increase until 6\% for short ones ($\lambda=30a$). The long wavelength approximation is responsible for the discrepancies obtained for thin non-uniform layers  (Fig.~\ref{alfven}c). Figure~\ref{alfven}d shows how  accurate Equation~(\ref{dampingrate}) is  for different combinations of wavelength, density contrast, and inhomogeneity length-scale. For thin layers ($l/a=0.1$) the inaccuracy of the long wavelength approximation produces differences up to $\sim$ 10\% for the combination of short wavelength with high contrast thread. For thick layers, 
differences of the order of 20\% are obtained. Here, the combination of large wavelength with high contrast thread produces the largest discrepancy. 

Numerical results allow the computation of more accurate values, but do no change the overall conclusion regarding the efficiency and properties of resonant damping of transverse oscillations in filament threads. Resonant damping in the Alfv\'en continuum appears as a very efficient mechanism for the explanation of the observed damping time scales.

\subsection{Resonant damping in the slow continuum}

Although the plasma $\beta$ in solar prominences is probably small, it is definitely nonzero. 
\cite{Soler09slowcont} have recently incorporated gas pressure to the cylindrical filament thread model of 
\cite{Arregui08thread}. This introduces the slow mode in addition to the kink mode and the 
Alfv\'en continuum. In the context of coronal loops, which are presumably hotter and denser than the surrounding corona, the ordering of sound, Alfv\'en and kink speeds does not allow for the simultaneous
matching of the kink frequency with both Alfv\'en and slow continua. Because of their relatively higher density and lower temperature conditions, this becomes possible in the case of filament threads. Therefore, the kink mode phase speed is also within the slow (or cusp) continuum, which extends between the internal and external sound speeds, in addition to the Alfv\'en continuum. This brings an additional damping mechanism. Its contribution to the total resonant 
damping has been assessed by \cite{Soler09slowcont}.

The study by \cite{Soler09slowcont} considers a filament thread model equivalent to that of \cite{Arregui08thread}, i.e., a straight cylinder with prominence-like conditions embedded in an unbounded corona, with a transverse inhomogeneous layer between both media. The same one-dimensional density profile given by Equation~(\ref{rhor}) is used. The internal and external densities are set to $\rho_i=5\times10^{-11}$ kg m$^{-3}$ and
$\rho_e=2.5\times10^{-13}$ kg m$^{-3}$, giving a density contrast of $\rho_{i}/\rho_e=200$. The plasma temperature is related to the density through the usual ideal gas equation and values of $T_i=8000$ K and $T_e=10^{6}$ K are taken. The magnetic field, pointing in the axial direction is taken to be uniform, with a value $B_0= 5$ G everywhere. Under this conditions, the plasma-$\beta\simeq0.04$.

In order to compute the analytical contribution to the damping of the kink mode due to the 
slow resonance  a similar method to the one outlined for the Alfv\'en resonance is followed. Assuming that the inhomogeneous transition region over which density varies is sufficiently small, compared to the tube radius, jump conditions can be used to obtain analytical expressions for the dispersion relation. The jump conditions for the Alfv\'en resonance are given in Equation~(\ref{jumpalfven}). The corresponding jump conditions for the slow resonance
at the point $r=r_S$ are

\begin{equation}
[\xi_r]=-\imath\pi\frac{k^2_z}{\omega^2_c|\partial_r\rho_0|_S}\left(\frac{c^2_s}{c^2_s+v^2_A}\right)^2\ p_T, \mbox{\hspace{1cm}} [p_T]=0, \mbox{\hspace{1cm}} \mbox{\rm at} \mbox{\hspace{1cm}} r=r_S,
\end{equation}

\noindent
where $\omega_c$  and $|\partial_r\rho_0|_s$ are the cusp frequency and the modulus of the radial derivative of the density profile, evaluated at the slow resonance and $c^2_s$ is the sound speed. 
Note that the factor $\left(\frac{c^2_s}{c^2_s+v^2_A}\right)$ is constant everywhere in the equilibrium. These jump conditions allow to obtain  a more general dispersion relation that now contains the contributions from both 
Alfv\'en and slow resonances as 

\begin{equation}
D(\omega,k_z)=-\imath\pi\frac{m^2/r^2_A}{\omega^2_A}\frac{\rho_i\rho_e}{|\partial_r\rho_0|_A}
-\imath\pi\frac{k^2_z}{\omega^2_c}\left(\frac{c^2_s}{c^2_s+v^2_A}\right)^2\frac{\rho_i\rho_e}{|\partial_r\rho_0|_S}.
\end{equation}

\noindent
To obtain an analytic expression for the damping rate of the kink mode the long-wavelength limit has to be considered. In terms of the physically relevant quantities, the damping time over the period can now be cast as

\begin{equation}
\frac{\tau_d}{P}=F\frac{1}{(l/a)}\left(\frac{\rho_i+\rho_e}{\rho_i-\rho_e}\right)\left[\frac{m}{\cos{\alpha_A}}+\frac{(k_za)^2}{m}\left(\frac{c^2_s}{c^2_s+v^2_A}\right)^2\frac{1}{\cos\alpha_S}\right]^{-1}.\label{dampingalfvenslow}
\end{equation}

\noindent
Here $F$ is the same numerical factor as in Equation~(\ref{dampingrate}), $\alpha_A=\pi(r_A-a)/l$, and $\alpha_S=\pi(r_S-a)/l$. The term with $k_z$ corresponds to the contribution of the slow resonance. If this term is dropped and $m=1$ and $\cos\alpha_A=1$ are taken Equation~(\ref{dampingalfvenslow}) becomes equivalent to Equation~(\ref{dampingrate}) which only takes into account the Alfv\'en resonance.

Equation~(\ref{dampingalfvenslow}) can now directly be applied to measure the relative contribution of each resonance to the total damping. To do that \cite{Soler09slowcont} assume $r_A\simeq r_S\simeq a$, for simplicity, so $\cos\alpha_A\simeq\cos\alpha_S\simeq 1$. The ratio of the two terms in Equation~(\ref{dampingalfvenslow}) is then

\begin{equation}\label{ratioas}
\frac{(\tau_d)_A}{(\tau_d)_S}\simeq\frac{(k_za)^2}{m^2}\left(\frac{c^2_s}{c^2_s+v^2_A}\right)^2.
\end{equation}

\noindent
A simple calculation shows that for the wavelength of observed filament thread oscillations, typically in between $10^{-3} < k_za < 10^{-1}$, the slow resonance is irrelevant to account for the kink mode damping. For instance, for $m=1$ and $k_z a=10^{-2}$ Equation~(\ref{ratioas}) gives $(\tau_d)_A/(\tau_d)_S\simeq 10^{-7}$. Even with the largest ratio that can be obtained, assuming a very large plasma-$\beta$, i.e., $c^2_s\rightarrow\infty$, so the factor $\left(\frac{c^2_s}{c^2_s+v^2_A}\right)\rightarrow 1$, gives a ratio of Alfv\'en to slow continuum damping times of $(\tau_d)_A/(\tau_d)_S\ll 1$.

\begin{figure}[!t]
\begin{minipage}{0.50\textwidth}
\includegraphics[width=1.0\textwidth]{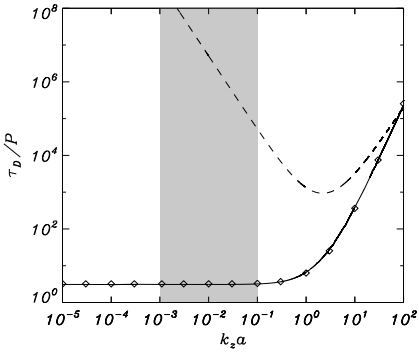}
\end{minipage}
\hspace{0.1cm}
\begin{minipage}{0.48\textwidth}
\caption{Ratio of the damping time to the period, $\tau_d/P$, as a function of the dimensionless wavenumber, $k_za$, corresponding to the kink mode for $l/a=0.2$. The solid line is the full numerical solution. The symbols and the dashed line are the results of the thin boundary approximation for the Alfv\'en and slow resonances. The shaded region represents the range of typically observed values for the wavelengths in prominence oscillations. Adapted from \citet{Soler09slowcont}. \label{resonantslow}}
\end{minipage}
\end{figure}

This  analytical result is further confirmed by \cite{Soler09slowcont} by performing  numerical computations outside the thin tube and thin boundary approximations, by solving the full resistive eigenvalue problem. Figure~\ref{resonantslow} displays the individual contribution of each resonance. The slow resonance  is seen to be much less efficient than the Alfv\'en resonance. For the wavenumbers relevant to observed prominence oscillations, the value of $\tau_d/P$ due to the slow resonance is in between 4 and 8 orders of magnitude larger than the same ratio obtained for the Alfv\'en resonance. On the other hand, the complete numerical solution (solid line) is close to the result for the Alfv\'en resonance. As seen in Sect.~ \ref{arregui-alfven} one can obtain values of $\tau_d/P\simeq 3$ in the relevant range of wavenumber. Note also that for short wavelengths ($k_za \geq 10^0$) the value of $\tau_d/P$ increases and the efficiency of the Alfv\'en resonance as a damping mechanism decreases. For the larger wavelengths considered to produce Figure~\ref{resonantslow}, $k_za \simeq 10^2$, both the slow and the Alfv\'en resonances produce similar (and inefficient) damping rates. The overall conclusion obtained by \cite{Soler09slowcont} is therefore that, although the plasma-$\beta$ in solar prominences is definitely non-zero, the slow resonance is very inefficient in damping the kink mode for typical prominence conditions and in the observed wavelength range. The damping times obtained with this mechanism are comparable to those due to the thermal effects discussed previously. Therefore, the resonant damping  of transverse thread oscillations is governed by the Alfv\'en resonance.

\section{Resonant absorption in a partially ionized filament thread}

From the results described so far resonant absorption in the Alfv\'en continuum seems to be the most efficient damping mechanism for the kink mode and the only one that can produce the observed damping time-scales. On the other hand, the effects of partial ionization could also be relevant, at least for short wavelengths.  The question arises on whether partial ionization affects the mechanism of resonant absorption.  \citet{Soler09rapi} have integrated both mechanisms in a non-uniform cylindrical filament thread model in order to assess both analytically and numerically the combined effects
of partial ionization and Alfv\'enic resonant absorption on the kink mode damping. Apart from the inherent relevance of this work in connection to the damping of prominence oscillations, this study constitutes the first of the kind that considers the resonant absorption phenomenon in a partially ionized plasma.

The filament thread model used by \citet{Soler09rapi} is an infinite straight cylinder with prominence-like conditions embedded in an unbounded coronal medium. Resonant damping is included in the model by connecting the prominence and coronal densities by means of a transitional layer, with a characteristic inhomogeneity length-scale $l$ (see expressions [\ref{rhor}] and [\ref{rhotrans}]). The plasma properties are now also characterized by the ionization fraction, $\mut_0$. The coronal plasma is assumed to be fully ionized, but  the prominence material is only partially ionized. As with the transverse density variation, the radial behavior of the ionization fraction in filament threads is unknown, but a one-dimensional transverse profile, similar to the one used to model the equilibrium density, can be assumed. The following profile for the ionization fraction, $\mut(r)$, is adopted

\begin{equation}
 \mut_{0}\left(r\right)=\left\{\begin{array}{clc}
 \mut_f,&{\rm if}&r\le a - l/2,   \\
 \mut_{\rm tr}\left(r\right),&{\rm if}&a-l/2<r<a+l/2,\\ 
 \mut_c,&{\rm if}&r\geq  a+l/2, \\
\end{array} \right. \label{eq:profilemu}
\end{equation}
with
\begin{equation}
 \mut_{\rm tr}\left(r\right)=\frac{\mut_f}{2}\left\{\left(1+\frac{\mut_c}{\mut_f}\right) - \left( 1-\frac{\mut_c}{\mut_f}\right)\sin \left[\frac{\pi}{l}\left( r-a\right)\right]\right\},
\end{equation}
where the filament ionization fraction, $\mut_f$, is considered a free parameter and the corona is assumed to be fully ionized, so $\mut_c = 0.5$. The non-uniform transitional layer of length $l$ therefore connects plasma with densities in the range $\rho_f$ and $\rho_c$ and ionization degrees in the range $\mut_f $ and $\mut_c$.
As before, the one-fluid approximation is used and for simplicity the $\beta=0$ limit, which excludes slow waves, is considered. The quantities $\eta$, $\etac$, and $\etah$ are now functions of the radial direction.

\subsection{Analytical results}

In order to obtain some analytical approximations, first, only Cowling's diffusion is considered. This basically introduces perpendicular magnetic diffusion and allows to derive an analytical dispersion relation for transverse oscillations, since the induction equation can be written in a compact form as follows

\begin{equation}
  \frac{\pd {\mathit {\bf B}}_1}{\pd t} = \frac{\Gamma^2_{\rm A} }{\va^2} \nabla \times \left( {\mathit {\bf v}}_1 \times {\mathit {\bf B}}_0\right), \label{eq:induction2}
\end{equation}

\noindent
where $\Gamma^2_{\rm A} \equiv \va^2 - i \omega \etac$ is the modified Alfv\'en speed squared \citep{Forteza08}. The dispersion relation for trapped waves is obtained by imposing that the solutions are regular at $r=0$, the continuity of  the radial displacement, $\xi_r$ , and the total pressure perturbation, $p_{\rm T}$, at the tube boundary, and that perturbations must vanish at infinity  \citep[see, e.g.,][]{ER83}. The dispersion relation reads $D\left( \omega, k_z \right) =0$, with

\begin{equation}
D\left( \omega, k_z \right) =  \frac{n_c}{\rho_c \left( \omega^2 -  k_z^2 \Gamma_{{\rm A}c}^2 \right)} \frac{K'_m \left( n_c a \right)}{K_m \left( n_c a \right)} - \frac{m_f}{\rho_f \left( \omega^2 - k_z^2 \Gamma_{{\rm A}f}^2 \right)} \frac{J'_m \left( m_f a \right)}{J_m \left( m_f a \right)}, \label{eq:dispernocapa}
\end{equation}

\noindent
where $J_m$ and $K_m$ are the Bessel function and the modified Bessel function of the first kind, respectively, and the quantities $m_f$ and $n_c$ are given by

\begin{equation}
 m_f^2 = \frac{\left(\omega^2 - k_z^2 \Gamma_{{\rm A}f}^2  \right)}{\Gamma_{{\rm A}f}^2},  \mbox{\hspace{0.5cm}} 
 n_c^2 = \frac{\left(k_z^2 \Gamma_{{\rm A}c}^2 -\omega^2 \right)}{\Gamma_{{\rm A}c}^2}.
\end{equation}

As shown in Section~\ref{resonantdamping}, an analytical dispersion relation in the case $l/a\neq0$ can be obtained in the
thin boundary approximation, by making use of the jump conditions for the radial displacement and the total pressure perturbation at the Alfv\'en resonance (expressions [\ref{jumpalfven}]). This dispersion relation is
\begin{equation}
D\left( \omega, k_z \right) =  -i \pi \frac{m^2 / r_{\rm A}^2}{\omega_k^2 \left| \partial_r \rho_0  \right|_{r_{\rm A}}}. \label{eq:dispercapa}
\end{equation}

\noindent
Note that Equation~(\ref{eq:dispercapa}) is formally identical to Equation~(\ref{disperalfven}), but now $D \left( \omega, k_z \right)$ is defined in Equation~(\ref{eq:dispernocapa}). By combining the thin tube and thin boundary approximations, and neglecting terms proportional to $k^2_z$, \citet{Soler09rapi} arrive at the following  short-hand expression for the damping time over the period  

\begin{equation}
 \frac{\td}{P} = \frac{2}{\pi} \left[ m \left( \frac{l}{a} \right) \left(  \frac{\rho_f - \rho_c}{\rho_f + \rho_c } \right) + \frac{2 \left(\rho_f \etactf + \rho_c \etactc \right)k_z a}{\sqrt{2 \rho_f \left(\rho_f + \rho_c \right) }} \right]^{-1}, \label{eq:tdp}
\end{equation}
with $\etactcf=\etac/v_{{\rm A}c,f}a$ the filament and coronal Cowling's diffusivities in dimensionless form. We see that the term related to the resonant damping is independent of the value of Cowling's diffusivity and, therefore, of the ionization degree, and takes the same form as in a fully ionized plasma, see e.g. Equation~(77) in \citet{goossens92} and Equation~(56) in \citet{RR02}. On the other hand, the second term related to the damping by Cowling's diffusion is proportional to $k_z$, so its influence in the long-wavelength limit is expected to be small. To perform a simple calculation, consider  the case  $m=1$, $k_z a = 10^{-2}$, and $l/a = 0.2$. This results in  $\tdp \approx 3.18$ for a fully ionized thread ($\mut_f = 0.5$), and  $\tdp \approx 3.16$ for an almost neutral thread ($\mut_f = 0.95$). The ratio $\tdp$ depends very slightly on the ionization degree, suggesting that resonant absorption dominates over Cowling's diffusion.

The relative importance of the two mechanisms can be estimated by taking the ratio of the two terms on the right-hand side of Equation~(\ref{eq:tdp})

\begin{equation}
 \frac{\left( \td \right)_{\rm RA}}{\left( \td \right)_{\rm C}} = \sqrt{\frac{2 \left(\rho_f + \rho_c \right)}{\rho_f}} \left( \frac{\rho_f \etactf + \rho_c \etactc}{\rho_f - \rho_c} \right) \frac{k_z a}{m \left( l/a \right)},
\end{equation}

\noindent
with $\left( \td \right)_{\rm RA}$ and $\left( \td \right)_{\rm C}$ the damping times due to resonant absorption and Cowling's diffusion, respectively. This last expression can be further simplified by considering that in filament threads $\rho_f \gg \rho_c$ and $\etactf \gg \etactc$, so that
\begin{equation}
 \frac{\left( \td \right)_{\rm RA}}{\left( \td \right)_{\rm C}} \approx \sqrt{2} \etactf \frac{k_z a}{m \left( l/a \right)}. \label{eq:ratiotaus}
\end{equation}
The efficiency of Cowling's diffusion with respect to that of resonant absorption increases with $k_z a$ and $\mut_f$ (through $\etactf$). Considering the same parameters as before, one obtains $\left( \td \right)_{\rm RA} / \left( \td \right)_{\rm C} \approx 2 \times 10^{-8}$ for $\mut_f = 0.5$, and $\left( \td \right)_{\rm RA} / \left( \td \right)_{\rm C} \approx 6 \times 10^{-3}$ for $\mut_f = 0.95$, meaning that resonant absorption is much more efficient than Cowling's diffusion. From Equation~(\ref{eq:ratiotaus}) it is also possible to estimate the wavenumber for which Cowling's diffusion becomes more important than resonant absorption, by setting $\left( \td \right)_{\rm RA} / \left( \td \right)_{\rm C} \approx 1$, as
\begin{equation}
 k_z a \approx \frac{m \left( l/a \right)}{\sqrt{2} \etactf}. \label{eq:kzCRA}
\end{equation}
Considering again the same parameters, Equation~(\ref{eq:kzCRA}) gives $k_z a \approx 5 \times 10^5$ for $\mut_f = 0.5$, and $k_z a \approx 1.7$ for $\mut_f = 0.95$. 
Note however that Equation~(\ref{eq:ratiotaus}) is only valid  for $k_z a \ll 1$, so resonant absorption is expected to be the dominant damping mechanism in the long-wavelength regime, even for an almost neutral filament plasma.

\subsection{Numerical results}

The analytical estimates described above are verified and extended by \citet{Soler09rapi} by numerically solving the full eigenvalue problem. Computations include now Hall's diffusion in addition to ohmic and Cowling's dissipation.

\begin{figure*}

  \includegraphics[width=0.5\textwidth]{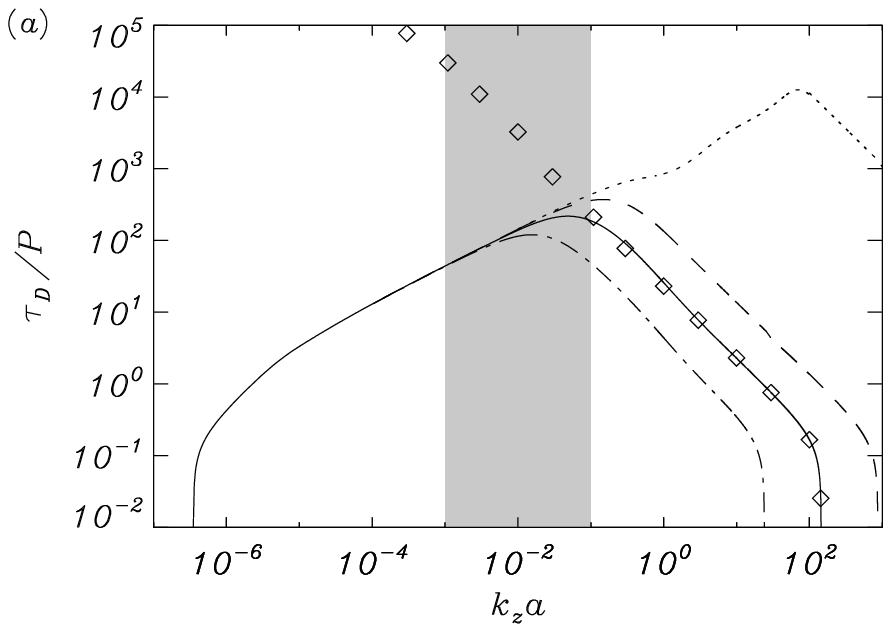}
  \includegraphics[width=0.5\textwidth]{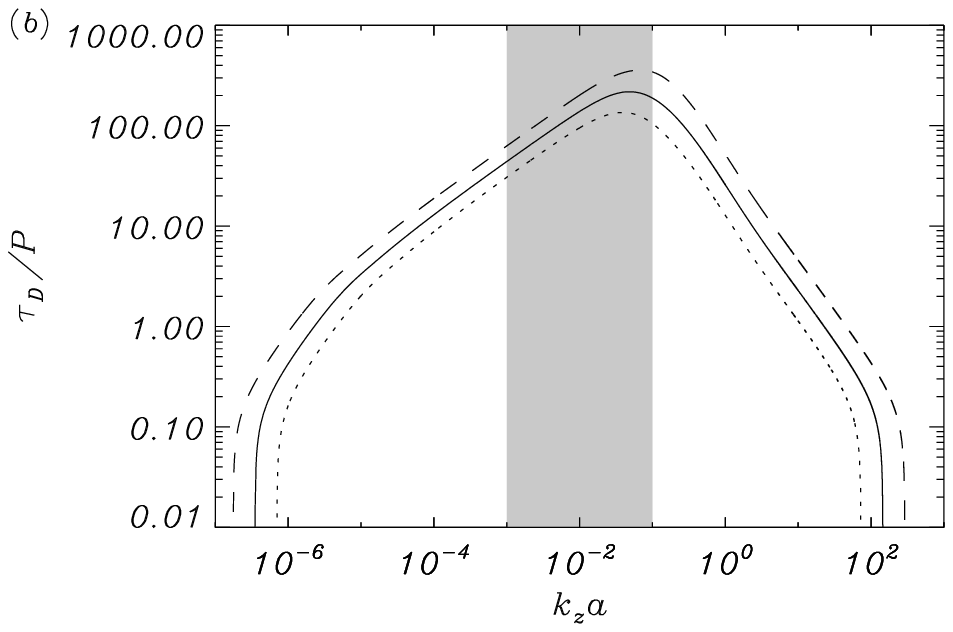}
\caption{Ratio of the damping time to the period of the kink mode as a function of $k_z a$ corresponding to a thread without transitional layer, i.e., $l/a=0$. ($a$) Results for $a = 100$~km considering different ionization degrees: $\tilde{\mu}_f = 0.5$ (dotted line),  $\tilde{\mu}_f = 0.6$ (dashed line), $\tilde{\mu}_f = 0.8$ (solid line), and $\tilde{\mu}_f = 0.95$ (dash-dotted line). Symbols are the approximate solution given by solving Equation~(\ref{eq:dispernocapa}) for $\tilde{\mu}_f = 0.8$. ($b$) Results for  $\tilde{\mu}_f = 0.8$ considering different thread widths: $a = 100$~km (solid line), $a = 50$~km (dotted line), and $a = 200$~km (dashed line). The shaded zone corresponds to the range of typically observed wavelengths of prominence oscillations. Adapted from \citet{Soler09rapi}.}
\label{fig:nocapa}       
\end{figure*}

First, a homogeneous filament thread without transitional layer ($l/a = 0$) is considered. This is the same case presented in Section~\ref{robertopicyl} with the addition of Hall's term in the induction equation.  Figure~\ref{fig:nocapa}a displays the obtained damping rate for different ionization degrees. In agreement with the results displayed for the kink mode in Figure~\ref{figpicyl1},  $\tdp$ has a maximum which corresponds to the transition between the ohmic-dominated regime, which is almost independent of the ionization degree, to the region where Cowling's diffusion is more relevant and the ionization degree has a significant influence. The approximate solution obtained by solving Equation~(\ref{eq:dispernocapa}) for a given value of  $\mut$  agrees very well with the numerical solution in the region where Cowling's diffusion dominates,  while it significantly diverges from the numerical solution in the region where ohmic diffusion is relevant. Within the range of typically reported wavelengths, $\tdp$ is between 1 and 2 orders of magnitude larger than the measured values,  so neither ohmic nor Cowling's diffusion can account for the observed damping time. Figure~\ref{fig:nocapa}b shows that the smaller the thread radius, the more rapidly the kink wave is attenuated. In addition, the critical wavenumbers are shifted when varying the thickness of the threads and the wavenumber range for which the kink wave propagates is wider for thick threads than for thin threads. The critical wavenumbers are far from the relevant values of $k_z a$ for the observed thread widths, and so they should not affect the kink wave propagation in realistic filament threads.

\begin{figure*}

  \includegraphics[width=0.5\textwidth]{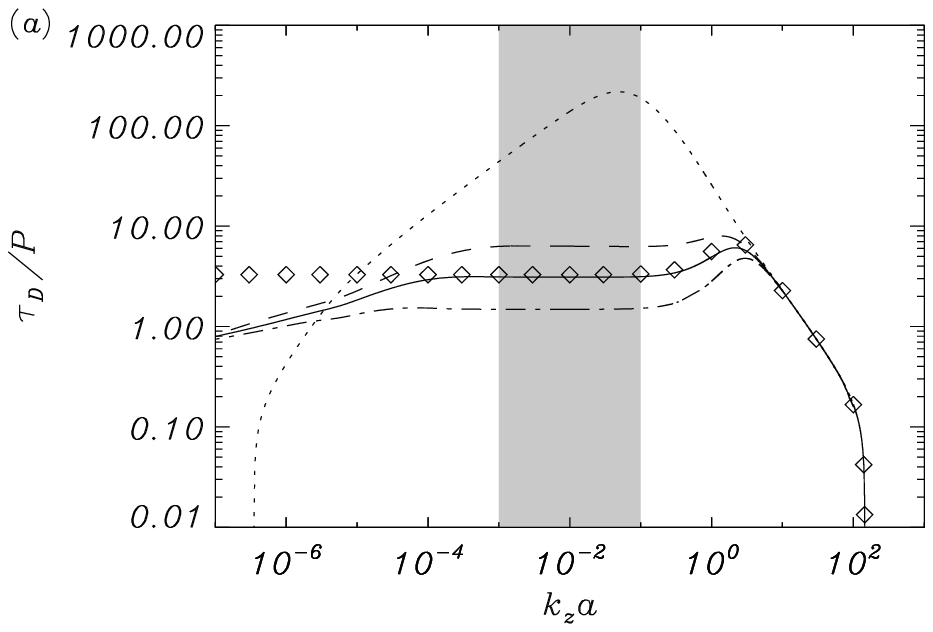}
  \includegraphics[width=0.5\textwidth]{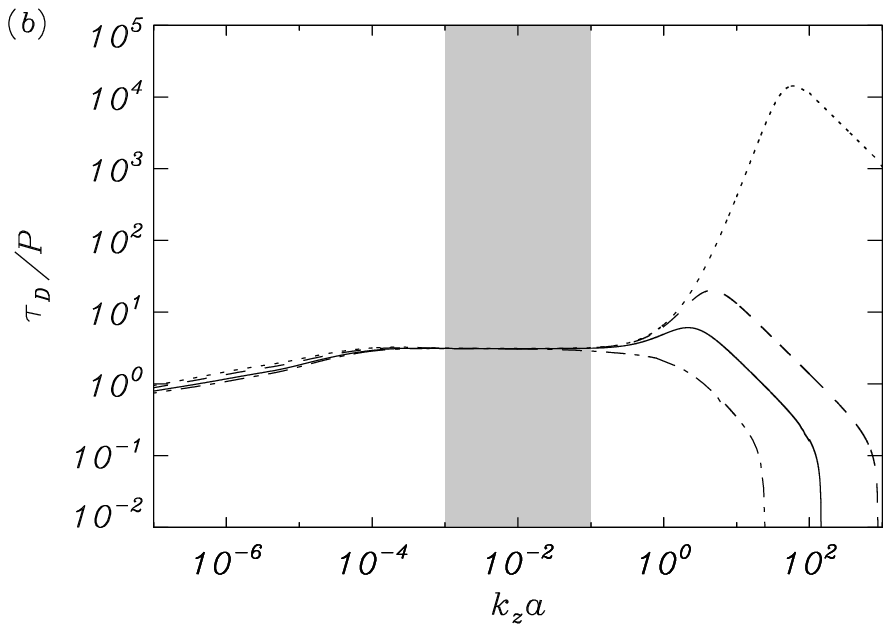}
\caption{Ratio of the damping time to the period of the kink mode as a function of $k_z a$ corresponding to a thread with an inhomogeneous transitional layer. ($a$) Results for $\mut_f = 0.8$ considering different transitional layer widths: $l/a = 0$ (dotted line),  $l/a = 0.1$ (dashed line), $l/a = 0.2$ (solid line), and $l/a  = 0.4$ (dash-dotted line). Symbols are the solution in the TB approximation given by solving Equation~(\ref{eq:dispercapa}) for $l/a = 0.2$. ($b$) Results for  $l/a = 0.2$ considering different ionization degrees: $\tilde{\mu}_f = 0.5$ (dotted line),  $\tilde{\mu}_f = 0.6$ (dashed line), $\tilde{\mu}_f = 0.8$ (solid line), and $\tilde{\mu}_f = 0.95$ (dash-dotted line). In both panels we have considered $a = 100$~km.  Adapted from \citet{Soler09rapi}.}
\label{fig:capa}       
\end{figure*}

Next, the inhomogeneous case ($l/a\neq0$) is analyzed, with the inclusion of resonant wave coupling in the Alfv\'en continuum. Figure~\ref{fig:capa}a displays some relevant differences with respect to the homogeneous case ($l/a = 0$). First,  the damping time is dramatically reduced for intermediate values of $k_z a$ including the region of typically observed wavelengths. In this region, the ratio $\tdp$ becomes smaller as $l/a$ is increased, a behavior consistent with damping by resonant absorption. The ratio $\tdp$ is independent of the thickness of the layer, and for large $k_z a$ coincides with the solution in the homogeneous case. The cause of this behavior is that perturbations are essentially confined within the homogeneous part of the thread for large $k_z a$, and therefore the kink mode is mainly governed by the internal plasma conditions. On the other hand, the solution for small $k_z a$ is completely different when $l/a \neq 0$. The inclusion of the inhomogeneous transitional layer removes the smaller critical wavenumber and consequently the kink mode exists for very small values of $k_z a$. Figure~\ref{fig:capa}a also shows a very good agreement between the numerical and the analytical solutions (obtained by solving the dispersion relation~[\ref{eq:dispercapa}]),  for wavenumbers above $k_z a \sim 10^{-4}$,  and a poor agreement in the range for which ohmic diffusion dominates, below $k_z a \sim 10^{-4}$. To understand this one has to bear in mind that dispersion relation~(\ref{eq:dispercapa}) takes into account the effects of resonant absorption and Cowling's diffusion, but not the influence of ohmic diffusion.
As Figure~\ref{fig:capa}b shows, the ionization degree is only relevant for large wavenumbers, where the damping rate significantly depends on the ionization fraction.

\begin{figure}
\begin{minipage}{0.50\textwidth}
  \includegraphics[width=1.0\textwidth]{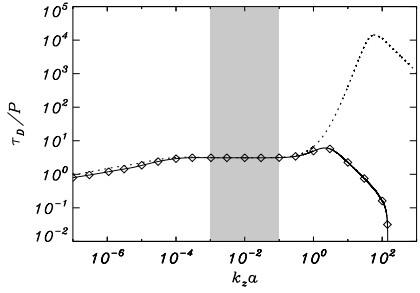}
  \end{minipage}
  \hspace{0.1cm}
  \begin{minipage}{0.48\textwidth}
\caption{Ratio of the damping time to the period of the kink mode as a function of $k_z a$ corresponding to a thread with $a = 100$ ~km and $l/a = 0.2$. The different linestyles represent the results for: partially ionized thread with $\mut_f = 0.8$ considering all the terms in the induction equation (solid line), partially ionized thread with $\mut_f = 0.8$ neglecting Hall's term (symbols), and fully ionized thread (dotted line).  Adapted from \citet{Soler09rapi}. \label{fig6soler} }
 \end{minipage}
\end{figure}

Finally, Figure~\ref{fig6soler} displays the ranges of $k_z a$ where Cowling's and Hall's diffusion dominate. As expected,  Hall's diffusion is irrelevant in the whole studied range of $k_z a$, while Cowling's diffusion dominates the damping for large $k_z a$. In the whole range of relevant wavelengths, resonant absorption is the most efficient damping mechanism, and the damping time is independent of the ionization degree as predicted by the analytical result  (Eq.~[\ref{eq:tdp}]). On the contrary,  ohmic diffusion dominates for very small $k_z a$. In that region,  the damping time related to Ohm's dissipation becomes even smaller than that due to resonant absorption, meaning that the kink wave is mainly damped by ohmic diffusion.

\section{Application to prominence seismology}

Solar atmospheric seismology aims to determine physical parameters in magnetic and plasma structures that are difficult to measure by direct means. It is a remote diagnostics method that combines observations of oscillations and waves in magnetic structures, together with theoretical results from the analysis of oscillatory properties of models of those structures. The philosophy behind this discipline is akin to that of Earth seismology, the sounding of the Earth interior using seismic waves, and helio-seismology, the acoustic diagnostic of the solar interior. It was first suggested by \cite{uchida70} and \cite{REB84}, in the coronal context, and by \cite {TH95} in a prominence context.  The last years increase in the number and quality of high resolution observations has lead to its rapid development. In the context of  coronal loop oscillations, recent applications of this technique have allowed the estimation and/or restriction of parameters such as the magnetic field strength \citep{Nakariakov01}, the Alfv\'en speed in coronal loops \citep{temury03,Arregui07,GABW08}, the transversal density structuring \citep{verwichte06}, or the coronal density scale height  \citep{AAG05,Verth08}. Its application to prominences is less developed. Some recent results of the MHD prominence seismology technique are shown.

\subsection{Seismology using the period of filament thread oscillations}

The first prominence seismology application using Hinode observations of flowing and transversely oscillating threads was presented by \cite{Terradas08hinode}, using observations obtained in an active region filament by \cite{Okamoto07}. The observations show a number of  threads that flow following a path parallel to the photosphere while they are oscillating in the vertical direction. \cite{Terradas08hinode} interpret these oscillations in terms of the kink mode of a magnetic flux tube. By using previous theoretical results from a normal mode analysis in a two-dimensional piecewise filament thread model by \cite{diaz02} and \cite{DR05}, these authors find that, although it is not possible to univocally determine the physical parameters of interest, a one-to-one relation between the thread Alfv\'en speed and the coronal Alfv\'en speed can be established.  This relation comes in the form of  a number of curves relating the two Alfv\'en speeds for different values of the length of the magnetic flux tube and the density contrast between the filament and coronal plasma.  The obtained curves have an asymptotic behavior for large values of the density contrast, typical of filament to coronal plasmas, and hence a lower limit for the thread Alfv\'en speed can be obtained. Further details on this study can be found in \cite{Terradas08hinode} and \cite{Oliver09}.

A recent application of the prominence seismology technique, using the period of observed filament thread transverse oscillations can be found in \citet{Lin09}. These authors find observational evidence about swaying motions of individual filament threads from high resolution observations obtained with the Swedish 1-m Solar Telescope in La Palma. The presence of waves propagating along individual threads was already evident in, e.g., \citet{Lin07}. However, the fact that line-of-sight oscillations are observed in prominences beyond the limb, as well as in filaments against the disk, suggests that the planes of the oscillation may acquire various orientations relative to the local solar reference system. For this reason, \citet{Lin09} combine  simultaneous recordings of motions in the line-of-sight and in the plane of the sky, which leads to information about the orientation of the oscillatory plane in each case. Periodic oscillatory signals are obtained in a number of threads, that are next fitted to sine curves, from which the period and the amplitude of the waves are derived. The presence of different cuts along the structures allow \cite{Lin09} to obtain the phase difference between the fitted curves, which can be used to measure the phase velocities of the waves. The overall periods and mean velocity amplitudes that are obtained correspond to short period, $P\sim$ 3.6 s, and small amplitude $\sim 2$ km s$^{-1}$ oscillations. The information obtained from these H$_\alpha$ filtergrams in the plane of the sky is combined with H$_\alpha$ Dopplergrams, which allow to detect oscillations in the line-of-sight direction.  By combining the observed oscillations in the two orthogonal directions the full vectors are derived, which show that the oscillatory planes are close to the vertical.

\begin{figure*}[!t]
  \includegraphics[width=0.5\textwidth]{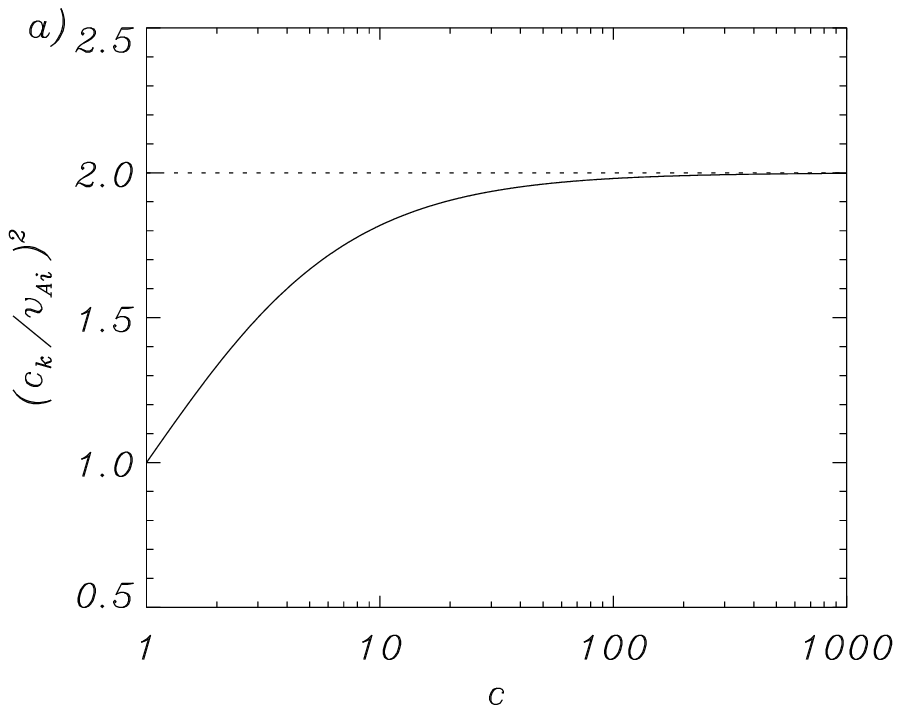}
  \includegraphics[width=0.5\textwidth]{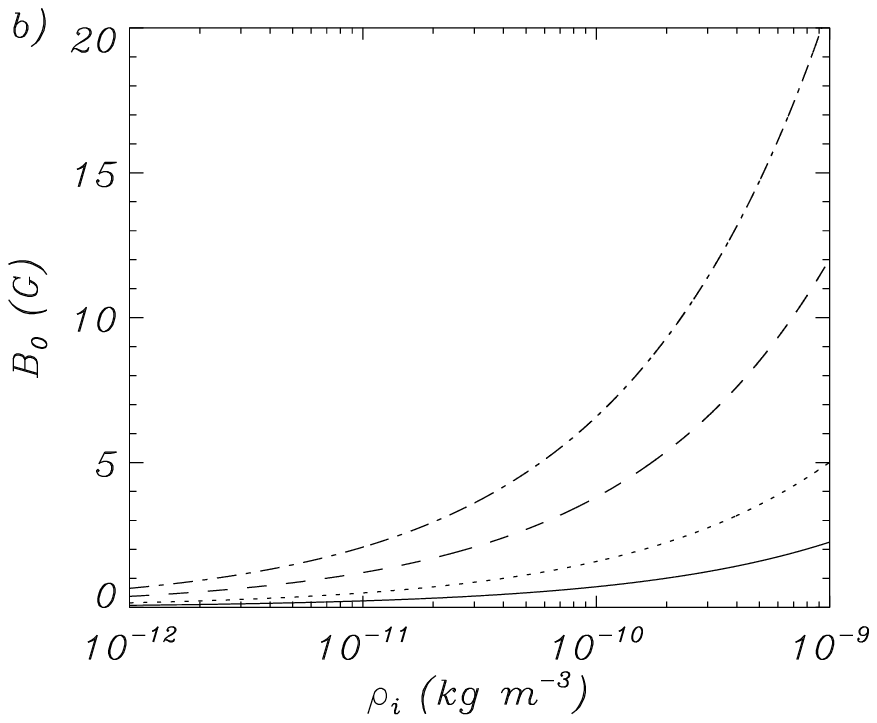}
\caption{{\em (a):} Ratio $c^2_k/v^2_{Ai}$ (solid line) as a function of the density contrast, $c$. The dotted line corresponds to the value of the ratio  $c^2_k/v^2_{Ai}$  for $c\rightarrow\infty$.
{\em (b):} Magnetic field strength as a function of the internal density, $\rho_i$, corresponding to some selected threads: thread 1 (solid line), thread 3 (dotted line), thread 5 (dashed line), and thread 7  (dash-dotted line). Adapted from \citet{Lin09}.}
\label{seismologylin09}       
\end{figure*}

\cite{Lin09} interpret the observed swaying thread oscillations as kink MHD waves supported by the thread body. By assuming the classic one-dimensional, straight, flux tube model a comparison between the observed wave properties and the theoretical prediction is performed on order to obtain the physical parameters of interest, namely the Alfv\'en speed and the magnetic field strength in the studied threads. To this end $c_k=\omega/k_z$ with $\omega$ defined by Equation~(\ref{kinkfrequency}) is used as an approximation to the kink speed, which can directly be associated to the measured phase velocity of the observed disturbances. Figure~\ref{seismologylin09}a shows that, for the large density contrasts expected for filament threads, the curve that relates the kink speed to the Alfv\'en speed becomes flat, and so the ratio is almost independent of the density contrast. This allows to simplify the kink speed to $c_k\simeq\sqrt{2}v_{Ai}$ and therefore obtain the thread Alfv\'en speed through $v_{Ai}\simeq V_{\rm phase}/\sqrt{2}$. The obtained values for a set of 10 threads  can be found in Table 2 in \cite{Lin09}. Once the Alfv\'en speed in each thread is determined, the magnetic field strength can be computed, if a given internal density is assumed. For a typical value of $\rho_i=5\times10^{-11}$ kg m$^{-3}$ magnetic field strengths in between 0.9 -3.5 G are obtained, for the analyzed events. If the unknown density is allowed  to vary in a range of plausible values, the derived magnetic field strength also changes accordingly, as can be seen in Figure~\ref{seismologylin09}b. The important conclusion that we extract  from the analysis by \cite{Lin09} is that prominence seismology is possible and works well, provided high resolution observations are available. The derived plasma parameters show rather different values for different threads, which indicates that the plasma parameters may be varying along individual threads  belonging to the same filament. This however does not come as a surprise in view of the highly inhomogeneous nature of these objects.

\begin{figure*}
  \includegraphics[width=0.5\textwidth]{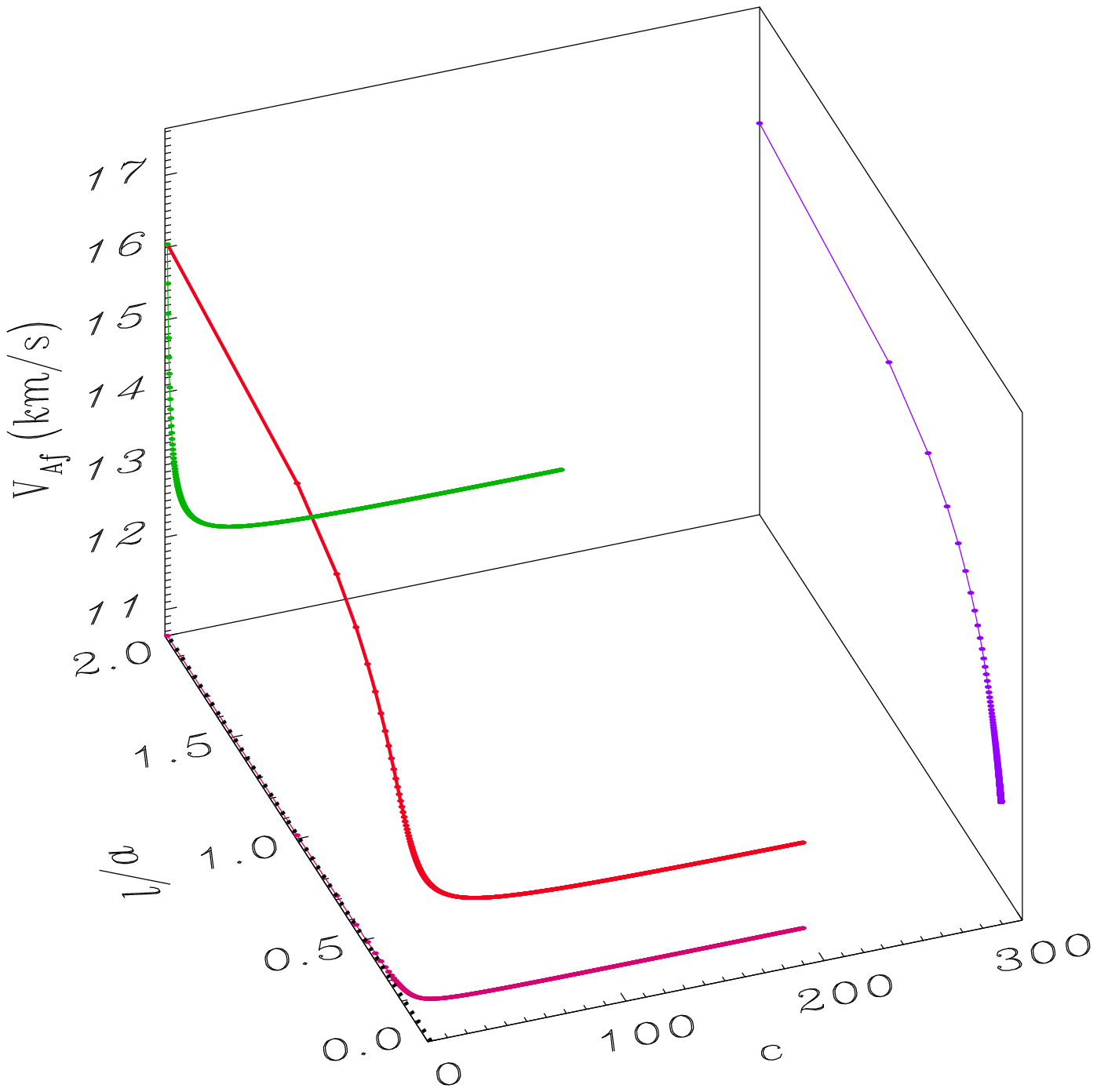}
  \includegraphics[width=0.5\textwidth]{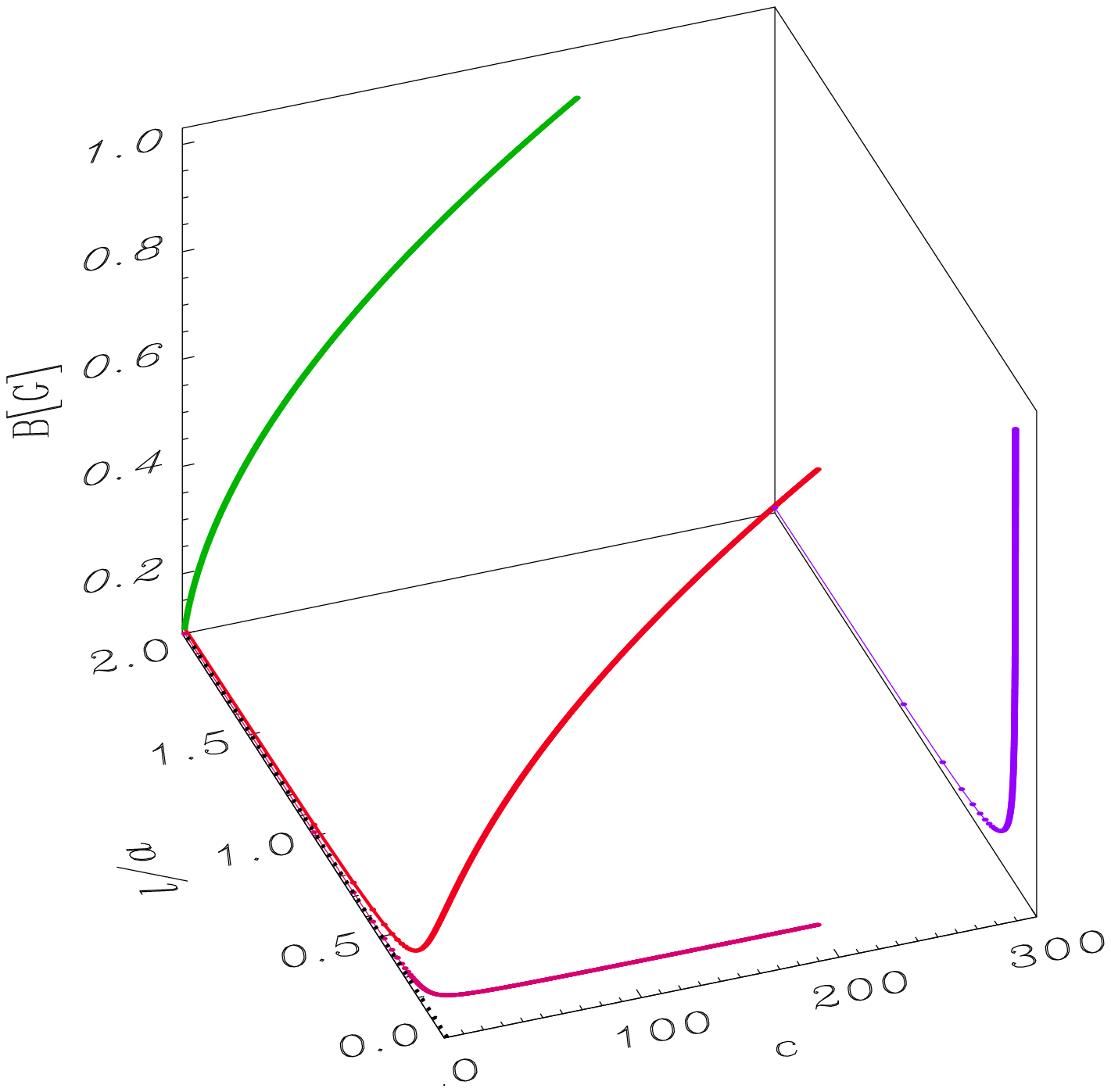}
\caption{{\em (Left):} Analytic inversion of physical parameters in the ($c$, $l/a$, $v_{Af}$) space for a filament thread oscillation with $P=3$ min, $\tau_d=9$ min, and a wavelength of $\lambda=3000$ km (see e.g. Lin et al. 2007). 
{\em (Right):} Magnetic field strength as a function of the density contrast and transverse inhomogeneity length-scale, derived from the analytic inversion for a coronal density of $\rho_c=2.5\times10^{-13}$ kg m$^{-3}$.}
\label{seismologyarregui}       
\end{figure*}

\subsection{Seismology using the period and damping of filament thread oscillations}

A feature clearly observed by \cite{Lin09} is that the amplitudes of the waves passing through two different cuts along a thread are notably different. Apparent changes can be due to damping of the waves in addition to the noise in the data.  Among the different damping mechanisms that are described in this paper, resonant absorption in the Alfv\'en continuum seems a very plausible mechanism and is open to direct application for filament thread seismology, using the damping as an additional source of information. This additional information has been used by  \cite{Arregui07,GABW08} in the context of transversely oscillating coronal loops and its application to filament threads is straightforward.

The analytical and numerical inversion schemes by \citet{Arregui07} and \citet{GABW08}
make use of the simple idea that it is the same magnetic structure, whose equilibrium conditions we are interested to assess, that is oscillating with a given period and undergoing a given damping rate.  By  computing the kink normal mode frequency and damping time as a function of the relevant equilibrium parameters for a one-dimensional model, the period and damping rate have the following dependencies

\begin{eqnarray}\label{relations}
P=P(k_z,c,l/a,\tau_{Ai}), \mbox{\hspace{1cm}} \frac{P}{\tau_d}=\frac{P}{\tau_d}(k_z,c,l/a),
\end{eqnarray}

\noindent
where $\tau_{Ai}$ the internal Alfv\'en travel time. In the case of coronal loop oscillations, an estimate for $k_z$ can be obtained directly from the length of the loop and the fact that the fundamental kink mode wavelength is twice this quantity. For filament threads, the wavelength of oscillations needs to be measured. Relations~(\ref{relations}) indicate that, if no assumption is made in any of the physical parameters of interest, we have two observed quantities, period and damping time, and three unknowns, density contrast, transverse inhomogeneity length-scale, and Alfv\'en travel time (or conversely Alfv\'en speed through the relation $v_{Ai}=L/\tau_{Ai}$). There are therefore infinite different equilibrium models that can equally well explain the observations.  These valid equilibrium models are displayed in Figure~\ref{seismologyarregui}a, where the analytical algebraic expressions in the thin tube and thin boundary approximations by \cite{GABW08} have been used to invert the problem.  It can be appreciated that, even if an infinite number of solutions is obtained, they define a rather constrained range of values for the thread Alfv\'en speed. Also, and because of the insensitiveness of the damping rate with density contrast, for the typically large values of this parameter in prominence plasmas,  the obtained solution curve displays an asymptotic behavior for large values of $c$. This allows us to obtain precise estimates for the filament thread Alfv\'en speed, $v_{Ai}\simeq 12$ km s$^{-1}$, and the transverse inhomogeneity length scale, $l/a\simeq 0.16$. Note that these asymptotic values can directly be obtained by inverting  Equations~(\ref{period}) and (\ref{dampingrate}) for the period and the damping rate in the limit $c\rightarrow\infty$.  The computation of the magnetic field strength from the obtained seismological curve requires the assumption of a particular value for either the filament or the coronal density. The resulting curve for a typical coronal density is shown in Figure~\ref{seismologyarregui}b. As can be seen, precise values of the magnetic field strength cannot be obtained, unless the density contrast is accurately known.

\section{Open Issues}

Solar prominences are, probably, the most complex structures in the solar corona. A full understanding of their formation, magnetic structure and disappearance has not been reached yet, and a lot of physical effects remain to be included in prominence models. For this reason, theoretical models set up to interpret small-amplitude oscillations and their damping are still poor. High-resolution observations of filaments suggest that they are made of threads whose thickness is at the limit of the available spatial resolution. Then, one may wonder whether future improvements of the spatial resolution will provide with thinner and thinner threads or, on the contrary, there is a lower limit for thickness and we will able to determine it in the near future.  The presence of these long and thin threads together with the place where they are anchored and the presence of flows along them suggest that they are thin flux tubes filled with continuous or discontinuous cool material. This cool material is probably subject to cooling, heating, ionization, recombination, motions, etc. which, altogether,  makes a theoretical treatment very difficult. For instance, in the case of the considered thermal mechanisms, up to now only optically thin radiation has been considered while, probably, the consideration of optically thick effects would be more realistic; the prominence heating mechanisms  usually taken into account are tentative and ``ad hoc''  while true prominence heating processes are still deeply unknown. An important step ahead would be to couple radiative transfer with magnetohydrodynamic waves as  a means to establish a relationship between velocity, density, magnetic field and temperature perturbations, and the observed signatures of oscillations like spectral line shift, width and intensity. Partial ionization is another topic of interest for prominence oscillations since, apart from influencing the behavior of magnetohydrodynamic waves, it poses an important problem for prominence equilibrium models  when gravity is taken into account and which is: How non-ionized prominence material is kept in the magnetic field?. Another issue which still remains a mystery is the triggering mechanism of small-amplitude oscillations. In the case of large-amplitude oscillations, observations provide with information about the exciting mechanism, however, the available observations of small-amplitude oscillations do not show any signature of the exciting mechanism. Are these oscillations of chromospheric or photospheric origin? Are they generated inside prominence magnetic structure by small reconnection events? Are they produced by weak external disturbances coming from far away in the solar atmosphere? The  changing physical conditions of prominence plasmas suggest that for an in-depth theoretical study of prominence oscillations more complex models together with numerical simulations are needed. Therefore, and as a step ahead, in the next future numerical studies of the time evolution of magnetohydrodynamic waves in partially ionized flowing inhomogeneous prominence plasmas, subject to different physical processes such as ionization, recombination, etc., should be undertaken. However, a full three-dimensional dynamical prominence model involving magnetic equilibrium, radiative transfer, etc. whose oscillatory behavior could be studied seems to be still far away in the future.

\begin{acknowledgements}
The authors acknowledge the financial support received from the Spanish MICINN, FEDER funds, and the Conselleria d'Economia, Hisenda i Innovaci\'o of the CAIB under Grants No. AYA2006-07637 and PCTIB-2005GC3-03. The authors also want to acknowledge the contributions to the research reported here made by M. Carbonell, A.J. D\'{\i}az, P. Forteza, R. Oliver, R. Soler, and J. Terradas.
\end{acknowledgements}



\end{document}